\documentclass[lettersize,journal]{IEEEtran}
\usepackage{cite}
\usepackage{amsmath,amssymb,amsfonts}
\usepackage{algorithmic}
\usepackage{textcomp}
\usepackage{enumerate}
\usepackage{booktabs}
\usepackage{amssymb}
\usepackage[ruled]{algorithm2e}
\usepackage{mathtools}
\usepackage{algorithmic}
\usepackage{array}
\usepackage{arydshln}
\usepackage[caption=true,font=normalsize,labelfont=sf,textfont=sf]{subfig}
\usepackage{color}
\usepackage{mathrsfs} 
\usepackage{multirow}
\usepackage{textcomp}
\usepackage{stfloats}
\usepackage{url}
\usepackage{verbatim}
\usepackage{graphicx}
\usepackage{subfig}

\usepackage{cite}
\usepackage{makecell}

\makeatletter
\renewcommand{\maketag@@@}[1]{\hbox{\m@th\normalsize\normalfont#1}}%
\makeatother
\begin{document}

\title{Edge-Fog Computing-Enabled EEG Data Compression via Asymmetrical Variational Discrete Cosine Transform Network}

\author{Xin Zhu,~\IEEEmembership{Student Member,~IEEE},
Hongyi Pan,~\IEEEmembership{Member,~IEEE},
        Ahmet Enis Cetin,~\IEEEmembership{Fellow,~IEEE}
\thanks{Xin Zhu and Ahmet Enis Cetin are affiliated with the
Department of Electrical and Computer Engineering, University of Illinois
Chicago, Chicago, IL 60607 USA. }
\thanks{Hongyi Pan is affiliated with the
Machine and Hybrid Intelligence Lab,  Northwestern University, Chicago, IL 60607 USA. }
\thanks{Corresponding author: Xin Zhu; e-mail: xzhu61@uic.edu}
}



\maketitle

\begin{abstract}
The large volume of electroencephalograph (EEG) data produced by brain-computer interface (BCI) systems presents challenges for rapid transmission over bandwidth-limited channels in Internet of Things (IoT) networks. 
To address the issue, we propose a novel multi-channel asymmetrical variational discrete cosine transform (DCT) network for EEG data compression within an edge-fog computing framework. At the edge level, low-complexity DCT compression units are designed using parallel trainable hard-thresholding and scaling operators to remove redundant data and extract the effective latent space representation. At the fog level, an adaptive filter bank is applied to merge important features from adjacent channels into each individual channel by leveraging inter-channel correlations. Then,  the inverse DCT reconstructed multi-head attention is developed to capture both local and global dependencies and reconstruct the original signals. 
Furthermore, by applying the principles of variational inference, 
a new evidence lower bound is formulated as the loss function, driving the model to balance compression efficiency and reconstruction accuracy.
Experimental results on two public datasets demonstrate that the proposed method achieves superior compression performance without sacrificing any useful information for BCI detection compared with state-of-the-art techniques, indicating a feasible solution for EEG data compression.

\end{abstract}


\section{Introduction}
%
%
%
%

\IEEEPARstart{T}he brain-computer interface (BCI) is emerging as a pivotal technology within the framework of Industry 4.0~\cite{douibi2021toward}. BCI facilitates direct communication between the human brain and machines, bypassing traditional physical interfaces~\cite{maiseli2023brain}. This capability holds transformative potential for enhancing automation and operational efficiency in industrial environments. For instance, BCI allows operators to control robotic arms or machinery with greater precision and less physical effort~\cite{an2024development}. Furthermore, BCI can enhance worker safety by monitoring cognitive load and stress levels~\cite{bagheri2022simultaneous}. 

\begin{figure}[htbp]
	\centering
		\includegraphics[scale=.4]{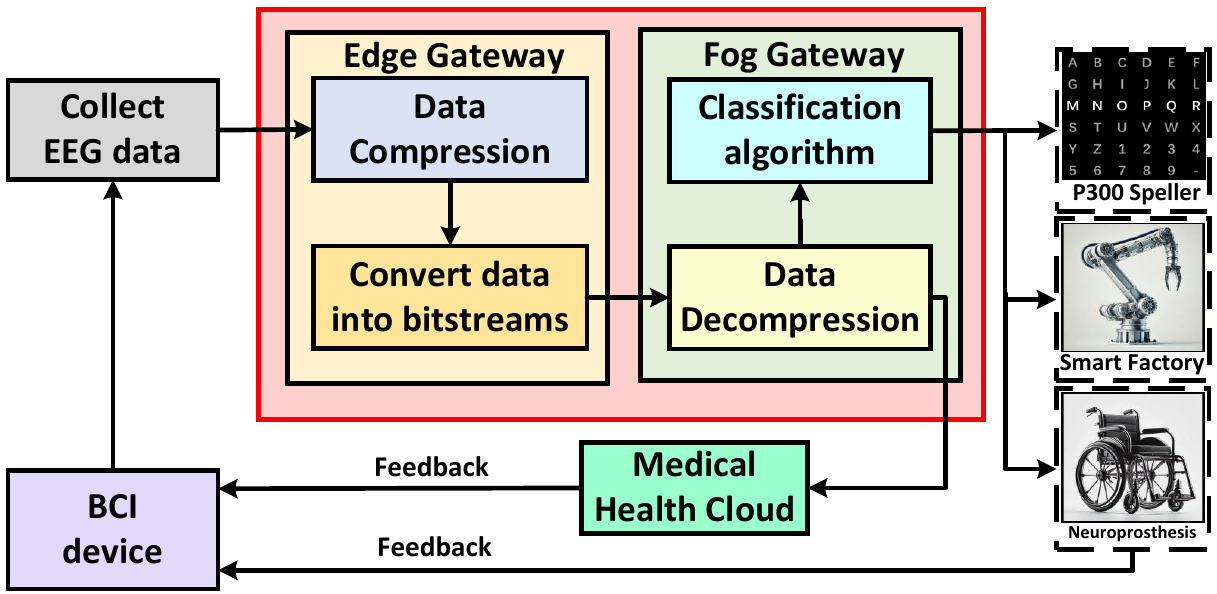}
	\caption{Edge-fog computing framework.}
	\label{Fig:background}
\end{figure}
The BCI system captures brain signals by measuring metabolic activity through functional near-infrared spectroscopy (fNIRS), magnetic activity using magnetoencephalography (MEG), or electrical activity via electroencephalography (EEG)~\cite{khan2021analysis}. EEG is recorded via electrodes placed on the scalp. It is widely utilized and well-suited for industrial applications due to its non-invasive nature, cost-effectiveness, high temporal resolution, and portability~\cite{altaheri2022physics}. 

The integration of BCI into the industrial field is intricately linked with the Internet of Things (IoT)~\cite{rodriguez2023fog,paszkiel2020using}. In smart factories, IoT enables various devices and systems to communicate and operate cohesively. BCI-equipped operators can manage multiple IoT-connected devices simultaneously to optimize workflows.
However, the vast volumes of EEG data generated by the BCI system pose significant challenges to rapid transmission and real-time analysis in bandwidth-limited channels~\cite{idrees2022edge}.
Edge and fog computing frameworks~\cite{hurbungs2021fog} address these challenges by deploying the edge gateway close to the data generators, such as biosensors, to perform real-time data compression, as illustrated in Fig.~\ref{Fig:background}. The compressed data is then transmitted to the fog gateway for decompression and feature extraction before being forwarded to BCI applications. Additionally, the reconstructed data is uploaded to the medical health cloud for storage and further analysis. Finally, both the applications and the medical cloud provide feedback to the users.
The fog gateway, positioned between the cloud and the sensors, provides an intermediary layer that enhances processing capabilities and reduces reliance on cloud resources for comprehensive data analysis. In contrast, while the edge gateway is situated near sensors, it is limited in its computational capabilities, rendering it insufficient for complex tasks such as training neural network models and performing extensive analyses on large-scale EEG datasets. Therefore, an efficient low-complexity EEG compression algorithm is required at the edge gateway to eliminate redundant data, reducing the latency, data transfer costs and bandwidth.

EEG signal compression techniques are generally divided into three main categories: traditional signal processing methods, neural network-based methods, and transform-based learning techniques~\cite{zhu2024electroencephalogram}.

Traditional signal processing methods, such as discrete cosine transform (DCT) and discrete wavelet transform (DWT), have been extensively employed in EEG data compression. These compression techniques can be categorized into single-channel and multi-channel schemes. In the single-channel methods, each EEG channel was processed independently using compression techniques, e.g., the fast DCT~\cite{birvinskas2015fast}, DWT~\cite{nguyen2018study}. In contrast, multi-channel compression algorithms perform simultaneous compression of EEG signals across all channels by exploiting inter-channel correlations. For instance, Hejrati \textit{et al.} utilized the K-means algorithm to cluster EEG channels~\cite{hejrati2017efficient}. Similar channels were then grouped to eliminate inter-channel dependencies, thereby reducing the data size.
 However, the maximum compression ratio achieved in~\cite{hejrati2017efficient} was smaller than 3, indicating a low compression efficiency.


Recently, neural networks have been increasingly utilized for data compression due to their capacity to learn complex data representations, such as the underlying patterns in EEG signals. Among these methods, autoencoders are the most frequently employed for compression tasks. For instance, the convolutional autoencoder (CAE)~\cite{al2018convolutional} and variational autoencoder~\cite{zancanaro2023veegnet} have demonstrated their efficacy in EEG data compression by applying convolutional layers, max-pooling layers, and evidence lower bound (ELBO) to optimize their models. Furthermore, Lerogeron $\textit{et. al}$ proposed a neural network-based approximation of dynamic time warping (DTW) as a reconstruction loss function, enhancing the ability of convolutional autoencoders to compress EEG signals while preserving essential information~\cite{lerogeron2023learning}. Nevertheless, these autoencoder-based models capture effective latent representations of EEG signals by incorporating numerous convolutional layers or linear layers, leading to increased computational complexity and higher hardware costs.

Over the past few decades, transform-based learning methods have gained considerable attention in EEG data compression for their ability to reduce data size while maintaining critical signal information. The authors in~\cite{hejrati2017new} proposed a learning-based adaptive transform method, which combines the DCT with an artificial neural network (ANN) to achieve near-lossless compression. The DCT is utilized to compress the original data, and then the MLP further compresses the primary DCT coefficients. Additionally, 
an asymmetrical sparse autoencoder~\cite{zhu2024electroencephalogram} was designed for compressing EEG sensor signals. This model integrates a low-complexity DCT layer within the encoder module to eliminate redundant coefficients and generate the latent representation. In contrast, the decoder module is equipped with additional linear layers to improve the quality of signal reconstruction. Nonetheless, These models function as single-channel approaches, where EEG signals are compressed and reconstructed on a per-channel basis, neglecting the correlations that exist among multi-channel EEG signals. As a result, the reconstruction accuracy is limited. 

To address the aforementioned issues, this paper proposes a novel multi-channel asymmetrical variational discrete cosine transform network (AVDCT-Net) to compress EEG signals across all channels simultaneously within the edge-fog computing framework. 
At the edge gateway, each channel is processed through a linear layer, followed by a low-complexity DCT compression unit (DCU) in the encoder module. 
In this unit, parallel hard-thresholding nonlinearities eliminate redundant data, while multiple scaling operators serve as an approximate filter bank within the DCT domain.
The entire compression process is guided by an ELBO-based cost function, enforcing a regularization constraint on the latent space coefficients to ensure a more structured and efficient representation. At the fog gateway, the decoder module reconstructs the EEG signal from compressed data utilizing an adaptive filter bank, inverse DCT (IDCT) reconstructed multi-head attention (IRMHA), and linear layers. The key contributions of this paper are summarized as follows: 

\begin{enumerate}[1)]
\item We propose a novel edge-fog computing-enabled multi-channel AVDCT-Net model to perform EEG data compression. By introducing the DCU to the encoder and employing the newly derived ELBO as an objective loss function, the AVDCT-Net achieves the trade-off between compression efficiency and reconstruction fidelity. Additionally, 
the encoder module exhibits low complexity, making it well-suited for deployment on edge gateways with limited computational resources.

\item An adaptive filter bank is implemented to promote the incorporation of important features from neighboring channels into each individual channel by leveraging the inter-channel correlations. This adaptive integration enhances reconstruction accuracy by compensating for any potential information loss that may occur during compression.
\item We design a new IRMHA based on the multi-head attention (MHA), hard-thresholding operator, and IDCT. The MHA mechanism strengthens the decoder’s feature extraction by capturing both local and global dependencies, while the IDCT aids in accurately reconstructing the original data.
\item Evaluation results demonstrate that the proposed method surpasses other state-of-art compression models in terms of compression efficiency and reconstruction accuracy on two EEG datasets: the BCI2 dataset~\cite{blankertz2004bci} and the BCI3 dataset~\cite{blankertz2006bci}. Moreover, it is experimentally shown that the proposed method achieves effective EEG signal compression without
compromising any important information for the BCI detection system.

\end{enumerate}  


\section{ Preliminaries}

\subsection{The Discrete Cosine Transform (DCT)}

The discrete cosine transform (DCT)~\cite{ahmed1974discrete} is extensively employed in data compression, particularly within established standards such as JPEG~\cite{wallace1992jpeg} for image compression and MPEG~\cite{le1991mpeg} for video compression.
Similar to the discrete Fourier transform, the DCT transforms a signal from the time domain to the frequency domain, allowing the data to be represented as a sum of cosine functions oscillating at various frequencies.
Given a sequence $\mathbf{x}$ of length $L$, the orthogonal type DCT-III is computed as follows:
\begin{equation}
    {X}_i=\sqrt{\frac{1}{L}}x_0 +\sqrt{\frac{2}{L}}\sum_{l=1}^{L-1}x_l\text{cos}\left[\frac{\pi}{L}\left(i+\frac{1}{2}\right)l\right],
\label{Eq:dct3}
\end{equation}
where $0\leq i \leq L-1$. The inverse DCT (IDCT) is defined as:
\begin{equation}
x_n = \alpha_n \sum_{i=0}^{L-1} X_i \cos\left[ \frac{\pi}{L} \left( i+ \frac{1}{2} \right) n \right], n = 0, \ldots, L-1,
\label{Eq:idct3}
\end{equation}
where $ \alpha_n=\sqrt{\frac{1}{L}} $ {if} $ n= 0$, otherwise $ \alpha_n=\sqrt{\frac{2}{L}}$.

The DCT is effective at concentrating the majority of signal energy into a limited number of low-frequency components. This capability allows for the preservation of critical information while enabling the elimination of less significant high-frequency data.

\subsection{The Variational Autoencoder (VAE)}
\label{sec: VAE}

A variational autoencoder (VAE)~\cite{chamain2022end} is a generative model that integrates principles from deep learning and Bayesian inference. 
Its application spans a wide range of tasks, including image synthesis, noise reduction and feature classification.
In the VAE formulation, the observed data \( \mathbf{x} \) are presumed to be generated by an underlying stochastic process, which involves an unobserved latent variable \(\mathbf{z} \) and is parameterized by $\vartheta$. Its marginal log-likelihood can be expressed as:
\begin{equation}
    \log p_{\vartheta}(\mathbf{x}) = \log \int p_{\vartheta}(\mathbf{x}, \mathbf{z}) \, d\mathbf{z}.
\label{eq: logp}
\end{equation}

Direct computation of this integral is often intractable due to the complexity of the joint distribution $p_{\vartheta}(\mathbf{x},\mathbf{z})$. To solve this problem, an approximate posterior distribution  \( q_{\varphi}(\mathbf{z}|\mathbf{x}) \) is introduced to serve as the \textbf{encoder}, where $\varphi$ represents the weights. After that, Eq.~(\ref{eq: logp}) is rewritten as:
\begin{equation}
\log p_{\vartheta}(\mathbf{x}) = \log \int q_{\varphi}(\mathbf{z}|\mathbf{x}) \frac{p_{\vartheta}(\mathbf{x},\mathbf{z})}{q_{\varphi}(\mathbf{z}|\mathbf{x})} \, d\mathbf{z}.
\label{eq: nlogp}
\end{equation}

Next, applying Jensen inequality to Eq.~(\ref{eq: nlogp}) yields the Evidence Lower Bound (ELBO) $\mathcal{L}(\mathbf{x})$:
\begin{equation}
\log p_{\vartheta}(\mathbf{x}) \geq \mathbb{E}_{q_{\varphi}(\mathbf{z}|\mathbf{x})} \left[ \log \frac{p_{\vartheta}(\mathbf{x}, \mathbf{z})}{q_{\varphi}(\mathbf{z}|\mathbf{x})} \right]=\mathcal{L}(\mathbf{x}).
\label{eq: ji}
\end{equation}

Given that $p_{\vartheta}(\mathbf{x},\mathbf{z})=p_{\vartheta}(\mathbf{x}|\mathbf{z}) p_{\vartheta}(\mathbf{z})$, the ELBO can be further decomposed as:
\begin{equation} 
\mathcal{L}(\mathbf{x}) = \mathbb{E}_{q_{\varphi}(\mathbf{z}|\mathbf{x})} \left[ \log p_{\vartheta}(\mathbf{x}|\mathbf{z}) \right] + \mathbb{E}_{q_{\varphi}(\mathbf{z}|\mathbf{x})} \left[ \log \frac{p_{\vartheta}(\mathbf{z})}{q_{\varphi}(\mathbf{z}|\mathbf{x})} \right],
\label{eq: CELBO}
\end{equation} 
where \( p_{\vartheta}(\mathbf{x}|\mathbf{z}) \) models the reconstruction \textbf{decoder} and its expected log-likelihood quantifies the reconstruction loss. Additionally, the second right-hand side term corresponds to the negative Kullback-Leibler divergence $- D_{KL}(q_{\varphi}(\mathbf{z}|\mathbf{x}) \| p_{\vartheta}(\mathbf{z}))$. It promotes a well-regularized latent space by penalizing deviations of the approximate posterior $q_{\varphi}(\mathbf{z}|\mathbf{x})$ from the prior distribution $p_{\vartheta}(\mathbf{z})$. Therefore, Eq.~(\ref{eq: CELBO}) can rewritten as:
\begin{equation} 
\mathcal{L}(\mathbf{x}) = \mathbb{E}_{q_{\varphi}(\mathbf{z}|\mathbf{x})} \left[ \log p_{\vartheta}(\mathbf{x}|\mathbf{z}) \right] - D_{KL}(q_{\varphi}(\mathbf{z}|\mathbf{x}) \| p_{\vartheta}(\mathbf{z})).
\label{eq: ELBO}
\end{equation} 
It is noticed that Eq.~(\ref{eq: ELBO}) is tractable and can be optimized efficiently. Hence, ELBO is a surrogate objective function when directly maximizing $\log p_{\vartheta}(\mathbf{x})$ is computationally prohibitive. 
Moreover, maximizing the ELBO involves two key objectives: improving data reconstruction accuracy via the expected log-likelihood term and regularizing the latent space through the KL divergence term.


\begin{figure*}[htbp]
	\centering
		\includegraphics[scale=.23]{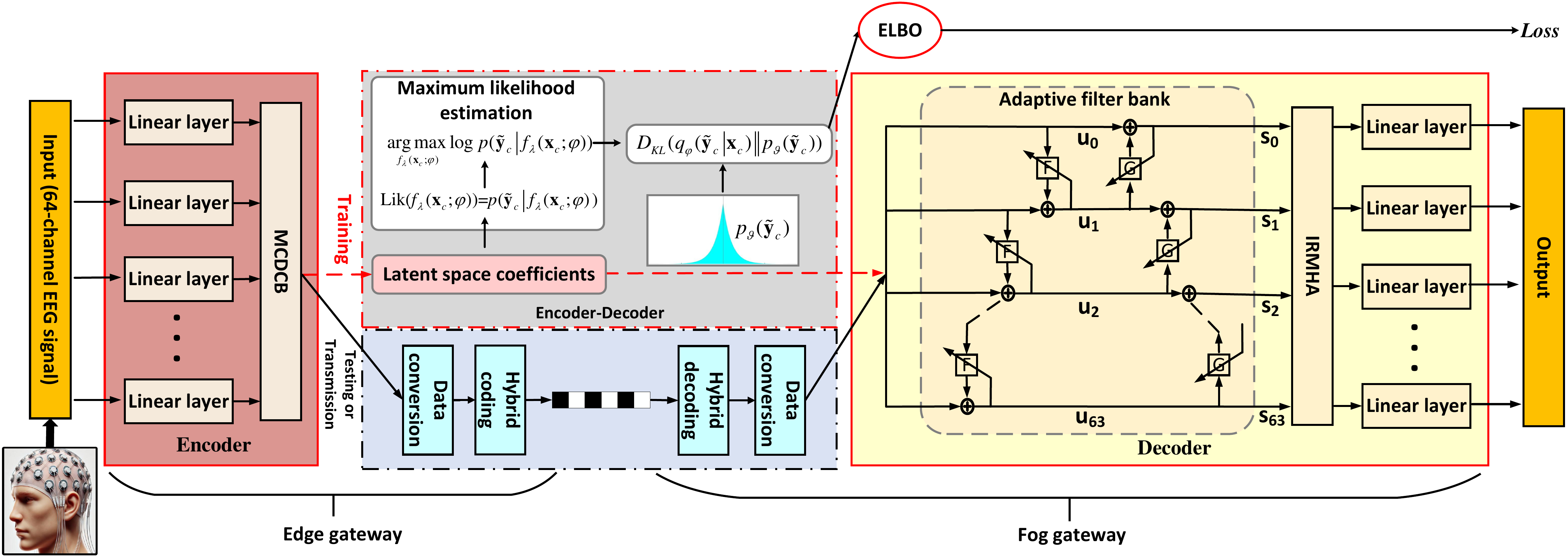}
	\caption{Block diagram of the asymmetrical variational discrete cosine transform network.}
	\label{Fig:AAEVDCT}
\end{figure*}

\section{Methodology}


This section presents the proposed multi-channel asymmetrical variational discrete cosine transform network (AVDCT-Net) for EEG data compression. 
The network employs an encoder with low computational complexity, making it suitable for deployment on a resource-constrained edge gateway. In contrast, the decoder is designed with a more intricate structure to improve the accuracy of data reconstruction. This asymmetry is feasible because the decoder runs on a more capable fog gateway, allowing for more intensive computational processes than the edge gateway~\cite{hurbungs2021fog}.


\subsection{Asymmetrical Variational Discrete Cosine Transform Network (AVDCT-Net)}


\begin{figure}[htbp]
	\centering
		\includegraphics[scale=.35]{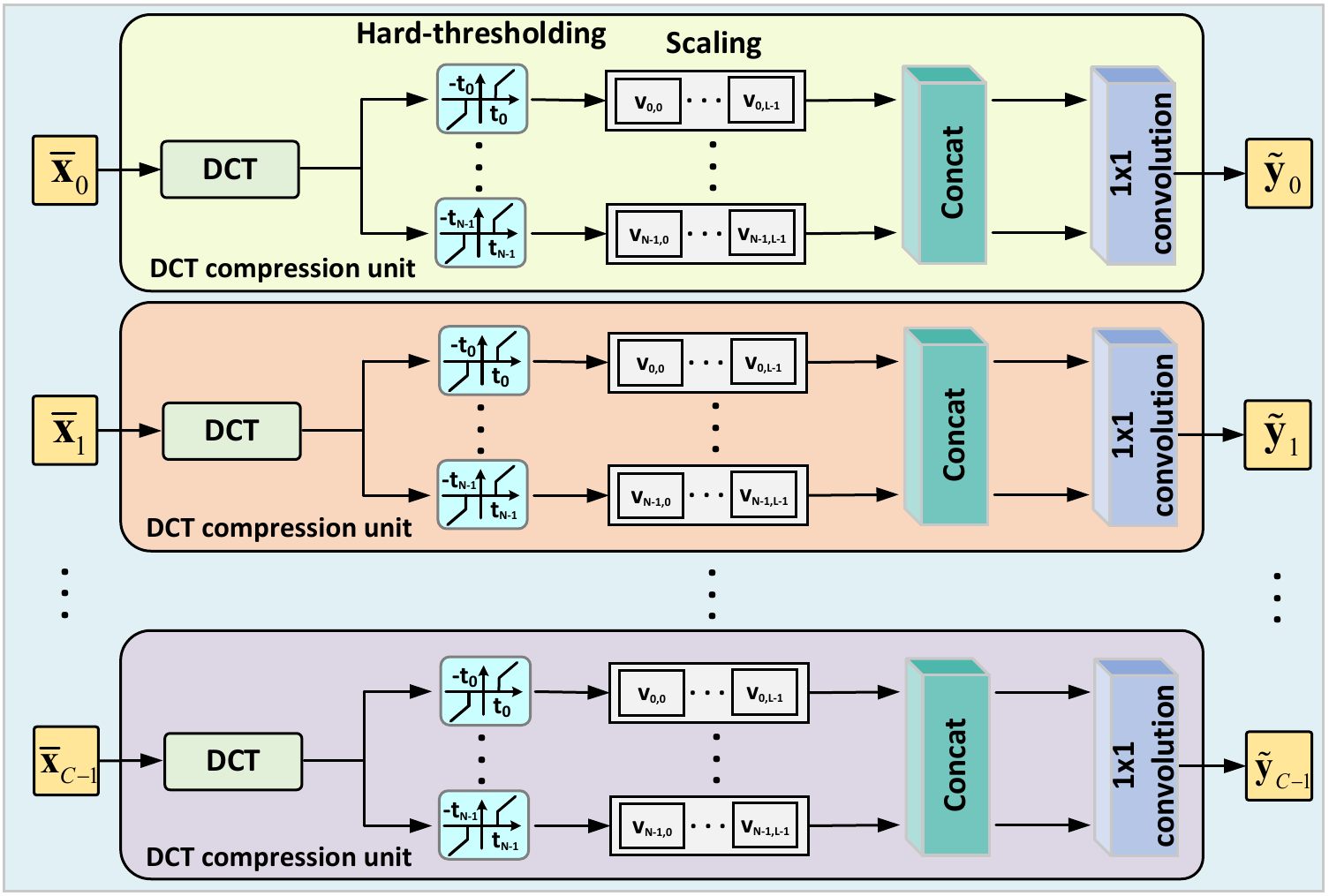}
	\caption{Multi-channel DCT compression block (MCDCB).}
	\label{Fig:MCDCU}
\end{figure}



In EEG systems, signals from all channels are captured simultaneously via scalp electrodes within the EEG headset. To maintain synchronization and minimize delays or data backlogs, AVDCT-Net leverages a multi-channel framework as shown in Fig.~\ref{Fig:AAEVDCT}, enabling the concurrent compression of signals across all channels.
In this study, we employ 64-channel EEG signals due to their widespread use in BCI experiments for classification and detection tasks~\cite{blankertz2006bci,blankertz2004bci,arican2019pairwise}. Additionally, the collected signal from each channel is segmented into short-time blocks of length $L$.

\subsubsection{Encoder module of the AVDCT-Net at the edge gateway}
As shown in Fig.~\ref{Fig:AAEVDCT}, the input block $\mathbf{x}_c\in\mathbb{R}^{L}$ is initially processed by the $c$-th linear layer to extract the features, producing the filtered output $\mathbf{\bar{x}}_c\in\mathbb{R}^{L}$, where $0\leq c\leq C-1$. Here, $C=64$ denotes the total number of channels. 
Next, a multi-channel DCT compression block is designed to perform EEG data compression, as shown in Fig.~\ref{Fig:MCDCU}. For each unit in the block, the DCT is employed to transform EEG signals from the time domain to the frequency domain using Eq.~(\ref{Eq:dct3}). The resulting transformed signal, $\hat{\mathbf{x}}_c\in\mathbb{R}^{L}$, is then passed through $N$ subbands. 
 Each subband is equipped with a trainable hard-thresholding operator and a scaling operator. 
 The hard-thresholding operator is applied to introduce nonlinearity between the DCT and scaling operator. Additionally, it can also eliminate small entries, which are typically redundant information and noise in the DCT domain. It is defined as:
\begin{equation}
    \begin{split}
        \tilde{\mathbf{x}}_{c,n}= \mathcal{R}_{c,n}\left(\hat{\mathbf{x}}_c\right)+\mathbf{t}_{c,n}\cdot\text{sign}\left(\mathcal{R}_{c,n}\left(\hat{\mathbf{x}}_c\right)\right),
    \end{split}
    \label{Eq:HT}
\end{equation}
where 
\begin{equation}
   \mathcal{R}_{c,n}\left(\hat{\mathbf{x}}_c\right)= \text{sign}\left(\hat{\mathbf{x}}_c\right)\cdot\left(|\hat{\mathbf{x}}_c|-\mathbf{t}_{c,n}\right)_{+}
\end{equation}
is the soft-thresholding operator~\cite{donoho1995noising}. $\mathbf{t}_{c,n}$ represents the threshold for $0\leq n \leq N-1$, which is trained using the back-propagation algorithm. $(\cdot)_+$ stands for the rectified linear unit (ReLU) function. While the soft-thresholding operator can be employed to denoise data, it may diminish the energy of prominent DCT coefficients, resulting in errors during data reconstruction. 

In contrast to the manually designed quantization matrix employed in the JPEG standards,  our model incorporates the trainable scaling vector that automatically optimizes the weighting of each DCT coefficient during the training process. The computation is defined as: 
\begin{eqnarray}\label{Eq:scaling}
{\mathbf{y}}_{c,n}=\tilde{\mathbf{x}}_{c,n}\circ {\mathbf{v}}_{c,n},
\end{eqnarray}
where $\circ$ stands for element-wise product. The scaling vector ${\mathbf{v}}_{c,n}\in\mathbb{R}^{L}$ is learned via the back-propagation algorithm.

Another perspective on the scaling layer pertains to the DCT convolution theorem i.e., symmetric convolution in the time domain can be achieved through element-wise multiplication in the DCT domain~\cite{martucci1994symmetric}. It is expressed as:
\begin{equation}\label{Eq:normal_dct_theorem}
    \mathbf{x} \otimes \mathbf{w} = {\mathcal{D}}^{-1}({\mathcal{D}}(\mathbf{x}\circ\mathbf{a}) \circ {\mathcal{D}}(\mathbf{w}\circ\mathbf{a}))\circ\mathbf{k},
\end{equation}
where $\mathbf{x}\in\mathbb{R}^L$ and $\mathbf{w}\in\mathbb{R}^L$ are two vectors.
$\mathcal{D}(\cdot)$ and $\mathcal{D}^{-1}(\cdot)$ denote the orthogonal type-III DCT and IDCT, respectively.
 $\otimes$ stands for the symmetric convolution~\cite{park2003m}. $\mathbf{a}$ represents a constant vector:
\begin{equation}\label{Eq:G}
\mathbf{a}[i]=\left\{
\begin{aligned}
&1/(2\sqrt{{L}}),  &i=0,\\
&\sqrt{1/(2L)} , &i > 0,\\
\end{aligned}
\right.
\end{equation}
where $0\leq i\leq L-1$. $\mathbf{k}[i]={1}/{\mathbf{a}[i]}$. Hence, each scaling layer in the DCT domain is akin to a trainable filter, as it can be translated into symmetric convolution within the time domain. Furthermore, the combination of $N$ scaling subbands constitutes a filter bank. It is experimentally shown in section~\ref{sec: Ablation experiments} that a combination filter bank leads to the improvement of feature extraction capability and compression performance. In addition, the proposed model achieves optimal compression results when $N=3$. 

After the scaling layer, the coefficients across all subbands are concatenated to form $\widehat{\mathbf{Y}}_{c}\in\mathbb{R}^{N\times L}$. Then, the one-by-one convolutional layer is introduced to reduce the dimensionality of $\widehat{\mathbf{Y}}_{c}$ from $\mathbb{R}^{N\times L}$ to $\mathbb{R}^{L}$, which increases the data compression efficiency in the latent space. Assume the convolutional layer output is $\tilde{\mathbf{y}}_{c}\in\mathbb{R}^{L}$.
The overall operation in each DCT compression unit is outlined in the Algorithm~\ref{algorithm:DCT}.
\begin{algorithm}[tb]
 \caption{DCT compression unit}
 \label{algorithm:DCT}
  \KwIn{ ${\mathbf{{x}}} \in\mathbb{R}^{L}$}
  \KwOut{ ${\mathbf{{z}}} \in\mathbb{R}^{L}$}
  \bf{Define:} $\mathbf{t}_n\in\mathbb{R}^{L}, \mathbf{V}_n\in\mathbb{R}^{L}, \mathbf{Conv}=\rm{Conv1D}(in=N, out=1,~kernel~size=1)$;\\
        ${\mathbf{y}}={\mathbf{DCT}}(\mathbf{x})\in\mathbb{R}^{L};$ \\
        \For{$n=0;n \le N-1;n=n+1$}
        {
            $\mathbf{e}_n= {\rm{sign}}\left(\mathbf{y}_n\right)\left(|\mathbf{y}_n|-\mathbf{t}_n\right)_{+}\in\mathbb{R}^{L};$\\
            $\mathbf{\tilde{e}}_n= \mathbf{e}_n+\mathbf{t}_n\circ{\rm{sign}}\left(\mathbf{e}_n\right)\in\mathbb{R}^{L};$\\
            $\mathbf{h}_n=\mathbf{\tilde{e}}_n\circ\mathbf{V}_n\in\mathbb{R}^{L};$\\
        }
        $\mathbf{h}={\rm{Concat}}(\mathbf{h}_n)\in\mathbb{R}^{N\times L};$ \\
        $\mathbf{z}={\mathbf{Conv}}(\mathbf{h})\in\mathbb{R}^{L};$ \\
        return ${\mathbf{z}}\in\mathbb{R}^{L};$ \\
\end{algorithm}

\begin{figure}[htbp]
	\centering
		\includegraphics[scale=.55]{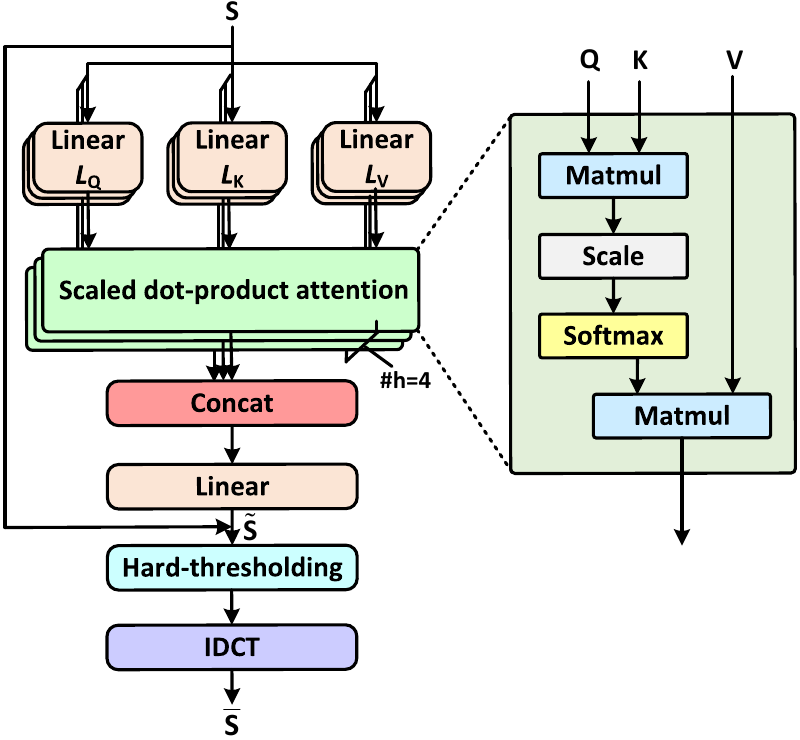}
	\caption{Inverse DCT reconstructed multi-head attention.}
	\label{Fig:IDRMA}
\end{figure}

\subsubsection{Decoder part of the AVDCT-Net  at the fog gateway}
After receiving the data from the encoder, the decoding process begins with the application of the adaptive filter bank. In multi-channel EEG systems, the signals recorded across different channels are often correlated due to the distributed nature of brain activity~\cite{dauwels2012near}. However, previous neural network-based compression models almost ignored such inter-channel correlations, which may provide important information for reconstruction. Therefore, the adaptive filter bank is employed to facilitate the integration of salient features from adjacent channels into each EEG channel. In this framework, the trainable scaling vectors $\mathbf{g}_{c}$ and $\mathbf{f}_{c}$ are employed as adaptive filters, based on the principle that scaling in the DCT domain can be equivalently transformed into convolution in the time domain. The computation is expressed as:
\begin{equation}
	\mathbf{u}_{c}= \begin{cases}
	      \tilde{\mathbf{y}}_{0}, &  \ c =0, \\
	      \tilde{\mathbf{y}}_{c}+\tilde{\mathbf{y}}_{c-1}\circ\mathbf{f}_{c-1}, &  \ 1 \leq c \leq C-1, 	
		   \end{cases}
\end{equation}
\begin{equation}
	{\mathbf{s}}_{c}= \begin{cases}
	      {\mathbf{u}}_{c}+{\mathbf{u}}_{c+1}\circ\mathbf{g}_{c}, &  \ 0 \leq c \leq C-2, \\
	      \mathbf{u}_c, &  \ c =C-1, 	
		   \end{cases}
\end{equation}
where $\tilde{\mathbf{y}}_{c}$ is the encoder output. $\mathbf{u}_{c}$ and ${\mathbf{s}}_{c}$ represent the outputs of the first and second filter banks, respectively, as shown in Fig~\ref{Fig:AAEVDCT}. During the lossy compression process, some critical information may be compromised. However, by employing an adaptive filter bank, each channel can recover potentially lost valuable information from correlated channels.


The adaptive filter bank output can be represented as a feature map $\mathbf{S}\in\mathbb{R}^{L\times C}$. 
In multi-channel EEG signals, spatially adjacent channels show inter-channel correlations, while each channel also has intra-channel correlations after applying the adaptive filter bank. Therefore,
the IDCT reconstructed multi-head attention (IRMHA) is designed to capture a comprehensive range of local and global dependencies from the feature map for original signal reconstruction. IRMHA comprises the multi-head attention (MHA), hard-thresholding operator, and IDCT.
Specifically, $\mathbf{S}$ is first processed through three separate linear transformations for each head. The transformations are described as:
\begin{equation}\label{Eq: WQKV}
    \mathbf{Q}_i = \mathbf{S} \mathbf{W}_i^Q+\mathbf{b}_Q, 
    \mathbf{K}_i = \mathbf{S} \mathbf{W}_i^K+\mathbf{b}_K, 
    \mathbf{V}_i = \mathbf{S} \mathbf{W}_i^V+\mathbf{b}_V,
\end{equation}
where $\mathbf{W}_i^Q\in\mathbb{R}^{C\times D}$, $\mathbf{W}_i^K\in\mathbb{R}^{C\times D}$, and $\mathbf{W}_i^V\in\mathbb{R}^{C\times D}$ stand for the transformation matrices that map the input dimensions to a lower-dimensional space specific to each head. $i$ indexes the head number. $D=\frac{C}{h}$. $h=4$ represents the number of heads. $\mathbf{b}_Q\in\mathbb{R}^{D}$, $\mathbf{b}_K\in\mathbb{R}^{D}$, and $\mathbf{b}_V\in\mathbb{R}^{D}$ denote trainable bias.
Then, each head applies the scaled dot-product attention mechanism independently:
\begin{equation}\label{Eq: attention}
\text{head}_i = \text{Attention}(\mathbf{Q}_i, \mathbf{K}_i,\mathbf{V}_i) = \text{softmax}\left( \frac{\mathbf{Q}_i \mathbf{K}_i^T}{\sqrt{D}} \right) \mathbf{V}_i.
\end{equation}

Next, the outputs of each head are concatenated and linearly transformed into the original input dimensionality:
\begin{equation}\label{Eq: multihead}
\mathbf{\widetilde{S}}= \text{Concat}(\text{head}_1, \text{head}_2, \ldots, \text{head}_{\beta}) \mathbf{W}^O,
\end{equation}
where $\mathbf{W}^O \in \mathbb{R}^{C \times C}$ is the transform weight matrix. $\mathbf{\widetilde{S}}\in\mathbb{R}^{L\times C}$ is MHA output. As the neural network deepens, it becomes increasingly susceptible to the vanishing gradient problem. To mitigate this issue, a residual connection is implemented, directly linking the input to the output of the MHA as shown in Fig.~\ref{Fig:IDRMA}. 
Subsequently, the hard-thresholding operator, as defined in Eq.~(\ref{Eq:HT}), is used to eliminate redundant DCT coefficients and introduce nonlinearity after the linear layer. 
The IDCT of the hard-thresholded output is then computed using Eq.~(\ref{Eq:idct3}), producing the result $\mathbf{\bar{S}}\in\mathbb{R}^{L\times C}$.
Finally,  the reconstructed $C$-channel EEG signal, represented as $\mathbf{Z}\in\mathbb{R}^{L\times C}$, is obtained using $C$ linear layers. 

\subsection{The Evidence Lower Bound (ELBO) for AVDCT-Net}

To optimize the compression ratio, it is a common practice to impose a sparsity constraint on the latent space coefficients~\cite{zhu2024electroencephalogram,zhu2024novel}.
While this approach mitigates redundancy, it also penalizes the primary coefficients, potentially leading to the loss of important information. Moreover, larger positive coefficients incur greater penalties according to the sparsity penalty definition~\cite{ng2011sparse}. 
To more effectively regularize the latent space and improve reconstruction accuracy, we derive a new ELBO for the multi-channel DCT compression block. 
In our model, Eq.~(\ref{eq: ELBO}) can be rewritten as follows:
\begin{equation} 
\mathcal{L}(\mathbf{x}_c) = \mathbb{E}_{q_{\varphi}(\tilde{\mathbf{y}}_{c}|\mathbf{x}_c)} \left[ \log p_{\vartheta}(\mathbf{x}_c|\tilde{\mathbf{y}}_{c}) \right] - D_{KL}(q_{\varphi}(\tilde{\mathbf{y}}_{c}|\mathbf{x}_c) \| p_{\vartheta}(\tilde{\mathbf{y}}_{c})),
\label{eq: NELBO}
\end{equation} 
where $q_{\varphi}(\tilde{\mathbf{y}}_{c}|\mathbf{x}_c)$ represents the approximate posterior distribution of the latent space DCT coefficients $\tilde{\mathbf{y}}_{c}$ conditioned on the observed data $\mathbf{x}_c$.
$p_{\vartheta}(\mathbf{x}_c|\tilde{\mathbf{y}}_{c})$ stands for the likelihood of reconstructing the original input $\mathbf{x}_c$ from its encoded representation $\tilde{\mathbf{y}}_{c}$. 
$p_{\vartheta}(\tilde{\mathbf{y}}_{c})$ is the prior distribution of $\tilde{\mathbf{y}}_{c}$. 

In our case, the actual distribution of $\tilde{\mathbf{y}}_{c}$ is unknown. However, the distribution of each DCT coefficient can be mathematically analyzed as a Laplacian distribution characterized by a location parameter $\mu=0$ and a scale parameter $\lambda$~\cite{lam2000mathematical}. 
Additionally, after the hard-thresholding and scaling operators, smaller, correlated variables are eliminated. 
The remaining larger variables represent distinct, essential signal components, which can be treated as independent. As a result, each latent space coefficient can be assumed to be independently and identically distributed in our model.
Hence, we represent  $q_{\varphi}(\tilde{{y}}_{c,l}|\mathbf{x}_c)$ using a Laplacian distribution, for $0\leq l \leq L-1$.
Furthermore, 
the probability density function (PDF) of $q_{\varphi}(\tilde{\mathbf{y}}_{c}|\mathbf{x}_c)$ can be defined as:
\begin{equation}
p(\tilde{\mathbf{y}}_{c}| f_{\lambda}({\mathbf{x}_c};\mathbf{\varphi})) =(\frac{1}{2f_{\lambda}({\mathbf{x}_c};\mathbf{\varphi})})^{L} \exp\left(-\frac{||\tilde{\mathbf{y}}_{c}||}{f_{\lambda}({\mathbf{x}_c};\mathbf{\varphi})}\right)
\label{Eq: px}
\end{equation}
where the scale function $f_{\lambda}({\mathbf{x}_c};\mathbf{\varphi})$ is parametrized by encoder neural network with parameters $\mathbf{\varphi}$. $||\cdot||$ denotes the 1-norm. 

\begin{figure}[tb]
	\centering
		\includegraphics[scale=.35]{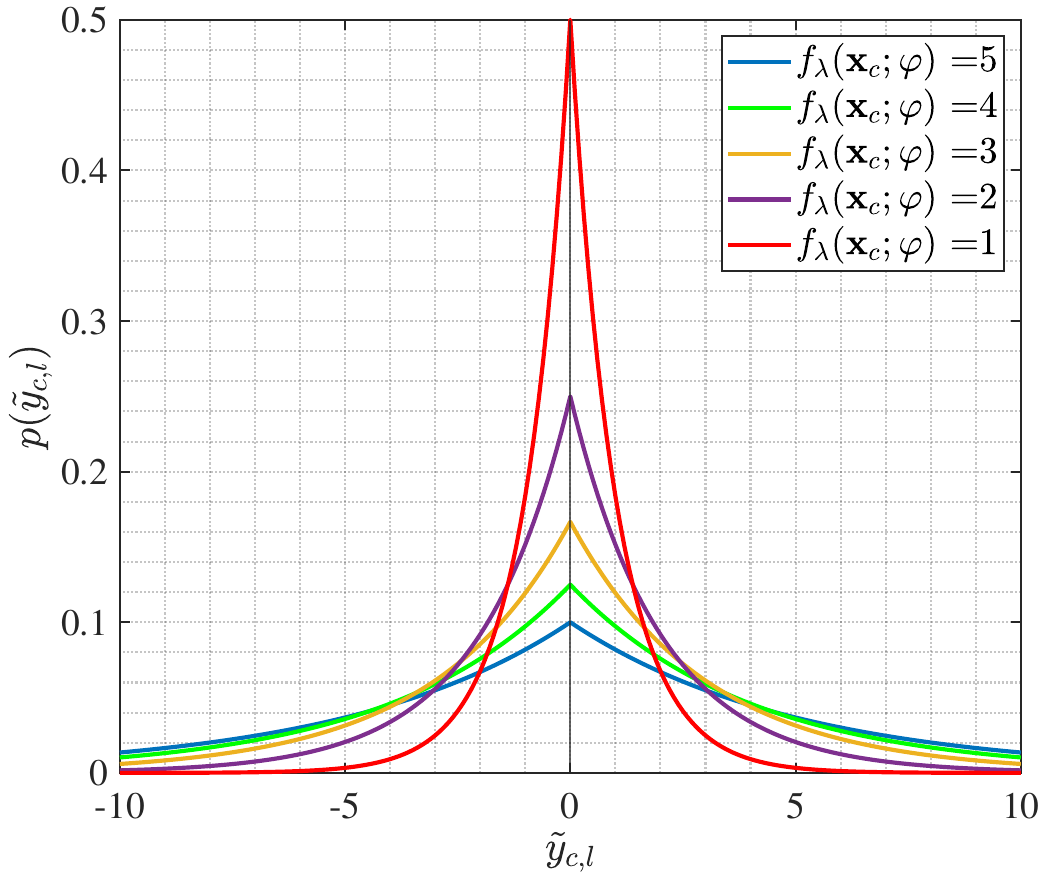}
	\caption{$p(\tilde{{y}}_{c,l})$ with five different $f_{\lambda}({\mathbf{x}_c};\mathbf{\varphi})$.}
	\label{Fig:LD}
\end{figure}

Assume $p(\tilde{{y}}_{c,l})$ is the PDF of $q_{\varphi}(\tilde{{y}}_{c,l}|\mathbf{x}_c)$.
Fig.~\ref{Fig:LD} illustrates the curves of $p(\tilde{{y}}_{c,l})$ corresponding to five different possible values of $f_{\lambda}({\mathbf{x}_c};\mathbf{\varphi})$. It is observed that when the value of $f_{\lambda}({\mathbf{x}_c};\mathbf{\varphi})$ decreases, the curve of $p(\tilde{{y}}_{c,l})$ becomes narrower and more peaked at zero, indicating a higher probability for zero. Additionally, the decay rate of the tails increases, resulting in a lower probability of extreme values. In EEG data compression, the promotion of sparsity within the latent space is pivotal for improving the compression ratio. Therefore, the value of $f_{\lambda}({\mathbf{x}_c};\mathbf{\varphi})$ is expected to be low. As mentioned the section~\ref{sec: VAE}, the prior distribution $p_{\vartheta}(\tilde{\mathbf{y}}_{c})$ acts as a regularizer by constraining $q_{\varphi}(\tilde{\mathbf{y}}_{c}|\mathbf{x}_c)$ to be close to a predefined distribution. Thus, we choose $p_{\vartheta}(\tilde{\mathbf{y}}_{c})$ as a Laplacian distribution with a small scale parameter $\lambda$. Its PDF is represented as:
\begin{equation}
p_{\vartheta}(\tilde{\mathbf{y}}_{c})=(\frac{1}{2\lambda})^{L} \exp\left(-\frac{||\tilde{\mathbf{y}}_{c}||}{\lambda}\right).
\label{Eq: ppx}
\end{equation}

After that, the term $D_{KL}(q_{\varphi}(\tilde{\mathbf{y}}_{c}|\mathbf{x}_c) \| p_{\vartheta}(\tilde{\mathbf{y}}_{c}))$ in Eq.~(\ref{eq: NELBO}) can be calculated as follows:
\begin{align}
&D_{KL}(q_{\varphi}(\tilde{\mathbf{y}}_{c}|\mathbf{x}_c) \| p_{\vartheta}(\tilde{\mathbf{y}}_{c}))\nonumber\\ &= \int q_{\varphi}(\tilde{\mathbf{y}}_{c}|\mathbf{x}_c) \log \frac{q_{\varphi}(\tilde{\mathbf{y}}_{c}|\mathbf{x}_c)}{p_{\vartheta}(\tilde{\mathbf{y}}_{c})} \, d\tilde{\mathbf{y}}_{c} \nonumber\\ 
&=\frac{ f_{\lambda}({\mathbf{x}_c};\mathbf{\varphi})L}{\lambda} + L\log \frac{\lambda}{f_{\lambda}({\mathbf{x}_c};\mathbf{\varphi})} - L.
\label{Eq: DKL}
\end{align}

However, the exact value of $f_{\lambda}({\mathbf{x}_c};\mathbf{\varphi})$ is still unknown. Next, we utilize maximum likelihood estimation to approximately solve $f_{\lambda}({\mathbf{x}_c};\mathbf{\varphi})$ based on the observed latent space coefficients $\tilde{\mathbf{y}}_{c}=[\tilde y_{c,1}, \ldots,\tilde y_{c,l}, \ldots, \tilde y_{c,L}]$. 
The likelihood of $f_{\lambda}({\mathbf{x}_c};\mathbf{\varphi})$ is computed as:
  \begin{align}\label{Eq:Lik}
   		{\rm {Lik}}(f_{\lambda}({\mathbf{x}_c};\mathbf{\varphi}))&=p({\tilde{\mathbf{y}}_{c}}|f_{\lambda}({\mathbf{x}_c};\mathbf{\varphi}))\nonumber\\
&= \prod_{l=0}^{L-1}\frac{1}{2f_{\lambda}({\mathbf{x}_c};\mathbf{\varphi})} \exp\left(-\frac{|\tilde{y}_{c,l}|}{f_{\lambda}({\mathbf{x}_c};\mathbf{\varphi})}\right).
  \end{align}


To find the maximum likelihood estimate, we solve the following optimization problem:
\begin{align}
\mathop{\arg\max}\limits_{f_{\lambda}({\mathbf{x}_c};\mathbf{\varphi})} \log p({\tilde{\mathbf{y}}_{c}}|f_{\lambda}({\mathbf{x}_c};\mathbf{\varphi}))
=\frac{1}{L}\sum\limits_{l=0}^{L-1}|\tilde y_{c,l}|.
\end{align}

Then, Eq~(\ref{Eq: DKL}) is rewritten as follows:
\begin{align}
&D_{KL}=\frac{\sum\limits_{l=0}^{L-1}|\tilde y_{c,l}|}{\lambda} + L\log{\lambda}-L\log{\frac{1}{L}\sum\limits_{l=0}^{L-1}|\tilde y_{c,l}|} - L.
\label{Eq: NDKL}
\end{align}
It is noticed that $D_{KL}$ achieves the minimum value when $\sum\limits_{l=0}^{L-1}|\tilde y_{c,l}|=\lambda L$. Unlike the approach in ~\cite{zhu2024electroencephalogram}, which penalizes each coefficient towards zero, our model constrains the sum of the latent space coefficients to a small nonzero value, $\lambda L$, thereby increasing the likelihood of retaining the important DCT coefficients while eliminating the redundant ones. As $\lambda$ approaches zero, $D_{KL}$ may exhibit numerical instability. To mitigate potential overflow, $D_{KL}(q_{\varphi}(\tilde{\mathbf{y}}_{c}|\mathbf{x}_c) \| p_{\vartheta}(\tilde{\mathbf{y}}_{c}))$ can also be replaced with $D_{KL}(p_{\vartheta}(\tilde{\mathbf{y}}_{c}) \| q_{\varphi}(\tilde{\mathbf{y}}_{c}|\mathbf{x}_c))$. This substitution helps prevent numerical issues and promote sparsity in the learned representations. 

As mentioned in section~\ref{sec: VAE}, the first right-hand term in Eq.~(\ref{eq: NELBO}) represents the reconstruction loss. Here we use mean square error as reconstruction loss, which is calculated as:
\begin{equation}
\mathbb{E}_{q_{\varphi}(\tilde{\mathbf{y}}_{c}|\mathbf{x}_c)} \left[ \log p_{\vartheta}(\mathbf{x}_c|\tilde{\mathbf{y}}_{c}) \right]=\frac{1}{L}\sum_{l=0}^{L-1} \left({x}_{c,l}-{z}_{c,l}\right)^2,
\label{eq: RL}
\end{equation}
where ${x}_{c,l}$ and ${z}_{c,l}$ represent the input and output element, respectively, $0\leq c \leq C-1$, $0\leq l \leq L-1$. 

Consequently, the new ELBO is described as:
\begin{align}
\mathcal{L}_n =-\frac{1}{L}\sum_{l=0}^{L-1} \left({x}_{c,l}-{z}_{c,l}\right)^2 -\varepsilon D_{KL},
\label{eq: FELBO}
\end{align}
where $\varepsilon$ is the weight of the KL-divergence term. The training objective is to maximize the ELBO. Additionally, we utilize a threshold parameter $\rho$ to adjust the compression ratio during the training stage.
When the percentage of zeros in $\tilde{\mathbf{y}}_{c}$ drops below the defined threshold $\rho$, the training procedure is stopped. In this process, small coefficients, considered redundant, are removed by being set to zero.


\subsection{Data Encoding and Storage}
\label{sec:transmission}
The data transmission part is comprised of data conversion and hybrid coding mechanisms, as depicted in Fig.~\ref{Fig:AAEVDCT}. 
The hybrid coding strategy is constructed by integrating Run-Length Encoding (RLE) \cite{akhter2010ecg} with the Lempel–Ziv–Markov chain algorithm (LZMA) \cite{tu2006novel} to enhance data compression efficiency.
Given that a double-precision floating-point number takes up 64 bits of memory, whereas an integer utilizes only 32 bits \cite{ratanaworabhan2006fast}, the conversion of floating-point numbers to integers is employed to optimize memory storage and enhance transmission rates. The conversion process is executed as follows:
\begin{eqnarray}\label{Eq:code}
\bar{\mathbf{y}}_{c}=\text{Round}(10^{\tau}\times {\tilde{\mathbf{y}}_{c}}/\omega) ,    
\end{eqnarray}
where $\tau$ is an integer, and $\omega$ is the scaling parameter utilized to adjust the compression ratio. $\text{Round}(\cdot)$ stands for the integer rounding function.
By utilizing RLE and LZMA compression techniques, the EEG signal is encoded into a bitstream, which is then prepared for wireless transmission.
The entire process begins with RLE, which efficiently removes sequential zero redundancies in the latent space. Subsequently, LZMA is applied to the RLE-processed data to achieve further compression.
Upon decoding, this compressed bitstream is reverted to its original integer form, after which the AVDCT-Net decoder reconstructs the signal on the fog gateway.

\section{Experimental Results}
\label{Experimental Results}

\subsection{Performance Metrics}
\label{sec:Metrics}
In this section, to evaluate the compression performances of different data compression algorithms, we analyze metrics such as compression ratio (CR), percent root mean square difference (PRD), normalized percent root mean square difference (PRDN), and quality score (QS)~\cite{zhu2024novel}. They are defined as follows:

\begin{itemize}

\item CR is determined by dividing the original data size $Q_r$ by the compressed data size $Q_c$, providing a crucial metric for evaluating the effectiveness of different algorithms in reducing data size:
\begin{equation}
{\rm{CR}}=\frac{Q_{r}}{Q_{c}}.
\end{equation}
A higher CR demonstrates that the model is more proficient at minimizing data size.

\item PRD evaluates the accuracy of data reconstruction in compression algorithms. It measures the discrepancy between the original data $e_m^o$ and the reconstructed data $e_m^r$, expressed as a percentage: 
\begin{equation}
{\rm{PRD}}=\sqrt{\frac{\sum_{m=1}^{M}\left(e_m^o-e_m^r\right)^2}{\sum_{m=1}^{M}\left(e_m^o\right)^2}}\times 100,
\label{Eq:prd}
\end{equation}
where $M$ represents the length of the data.
 A lower PRD value indicates that the reconstructed data closely matches the original data, signifying higher fidelity.

\item PRDN represents the normalized form of PRD. It is defined as:
\begin{equation}
{\rm{PRDN}}=\sqrt{\frac{\sum_{m=1}^{M}\left(e_m^o-e_m^r\right)^2}{\sum_{m=1}^{M}\left(e_m^o-\bar{e}\right)^2}}\times 100,
\end{equation}
where $\bar{e}$ denotes the mean of the original signal. PRDN provides a robust measure of similarity that is less sensitive to mean shifts in the original signal than PRD. A lower PRDN value signifies superior model performance.

\item QS assesses the overall performance of data compression algorithms. It integrates multiple performance aspects, such as compression ratio and reconstruction accuracy:
\begin{equation}
\rm{QS}=\frac{CR}{PRD}.
\end{equation}
A higher QS reflects that the algorithm excels in effectively reducing data size while maintaining high fidelity.
\end{itemize}

\subsection{Selected Dataset and Comparison Approaches}

In this study, the performances of different compression algorithms are evaluated on the BCI2~\cite{blankertz2004bci} and BCI3~\cite{blankertz2006bci} datasets:
1) The BCI2 (Dataset-IIa) dataset provides the EEG recordings for 10 30-minute sessions from 3 subjects (A/B/C). They were collected from 64 electrodes (64-channel) placed on the scalp with a sampling frequency of 160 Hz. To evaluate the compression performances of different algorithms, we utilize these 10 sessions from subject A, labeled sequentially from ``AA001'' to ``AA010''.  
2) The BCI3 (Dataset-II) dataset was recorded using Farwell and Donchin’s interface. It is composed of 4 datasets: The recordings designated as ``Subject\_A\_Train'' (ATr) and ``Subject\_A\_Test'' (ATe) were acquired from subject A, whereas the recordings labeled ``Subject\_B\_Train'' (BTr) and ``Subject\_B\_Test'' (BTe) were obtained from subject B. The training and testing datasets comprise 85 and 100 64-channel EEG recordings, respectively. Each recording comprises 15 rounds of repeated stimuli. In each round, the intensification period was 100 milliseconds (ms), and the inter-stimulus interval was 75 ms. Therefore the passing time for one recording is calculated as (100+75)$\times$12=2,100 ms. With 15 repetitions, the total time amounts to 31,500 ms. The sampling frequency on the BCI3 dataset is 240 Hz. Hence, each recording contains 7,560 data points for a single channel.


To verify the efficiency of the proposed data compression algorithm, we conduct a comparison with the asymmetrical sparse autoencoder with a DCT layer (ASAEDCT)~\cite{zhu2024electroencephalogram}, which has demonstrated excellent compression performance.
Besides, a state-of-the-art EEG compression method based on an Edge-Fog computing architecture is selected as a benchmark for comparison~\cite{idrees2022edge}. It is developed using K-means clustering and Huffman encoding (KCHE). To enhance the compression ratio, we refine KCHE by converting floating-point numbers into integers prior to transmission, as described in Eq.~(\ref{Eq:code}). 
In addition, we utilize the DCT-Perceptron~\cite{pan2024multichannel} and VAE~\cite{oliveira2023early} for further comparison, given that they employ low-complexity encoders. 

\subsection{Implementation Details}

The proposed model is implemented in Python on a PC equipped with an Intel Core i7-12700H CPU (4.7 GHz) and 16 GB of RAM.
The AdamW optimizer~\cite{loshchilov2017decoupled} is applied to train the model. To achieve a tradeoff between compression ratio and reconstruction accuracy, $\lambda$ and $\rho$ are chosen as 1e-5 and 0.6, respectively. In addition, the batch size and the learning rate are set as 16 and 0.001, respectively. Moreover, to reduce the trainable parameters and computation costs, we select a small block size $L=64$ and all channels use the same training parameters by adopting parameter sharing~\cite{christianos2021scaling}. 
\begin{table*}[htbp]
\caption{Compression experiment on the BCI2 dataset.}
\label{tab: Compression experiment on BCI2 dataset}
    \centering
    \scalebox{0.85}{
\renewcommand{\arraystretch}{1.5}
\begin{tabular}{c:cccc:cccc:cccc:cccc:cccc}
\Xhline{0.75pt}
\multicolumn{1}{c}{\multirow{2}{*}{\textbf{Sub}}}
&\multicolumn{4}{c}{\textbf{ASAEDCT}~\cite{zhu2024electroencephalogram}} 
&\multicolumn{4}{c}{\textbf{KCHE}~\cite{idrees2022edge}}
&\multicolumn{4}{c}{\textbf{DCT-Perceptron}~\cite{pan2024multichannel}}
&\multicolumn{4}{c}{\textbf{VAE}~\cite{oliveira2023early}}
&\multicolumn{4}{c}{\textbf{AVDCT-Net }(Proposed)}\\
\Xcline{2-21}{0.75pt}
\multicolumn{1}{c}{\multirow{2}{*}{{}}}
&
\multicolumn{1}{c}{\textbf{CR}}&\multicolumn{1}{c}{\textbf{PRD}} &\multicolumn{1}{c}{\textbf{PRDN}} &\multicolumn{1}{c}{\textbf{QS}}  
&\multicolumn{1}{c}{\textbf{CR}}&\multicolumn{1}{c}{\textbf{PRD}} &\multicolumn{1}{c}{\textbf{PRDN}} &\multicolumn{1}{c}{\textbf{QS}}
&\multicolumn{1}{c}{\textbf{CR}}&\multicolumn{1}{c}{\textbf{PRD}} &\multicolumn{1}{c}{\textbf{PRDN}} &\multicolumn{1}{c}{\textbf{QS}}
&\multicolumn{1}{c}{\textbf{CR}}&\multicolumn{1}{c}{\textbf{PRD}} &\multicolumn{1}{c}{\textbf{PRDN}} &\multicolumn{1}{c}{\textbf{QS}}
&\multicolumn{1}{c}{\textbf{CR}}&\multicolumn{1}{c}{\textbf{PRD}} &\multicolumn{1}{c}{\textbf{PRDN}} &\multicolumn{1}{c}{\textbf{QS}}
 \\
\Xhline{0.75pt}
1 
&{5.91} & {16.62} & {16.67} & {0.36} 
&{3.73} & {16.46} & {16.50} & {0.23}
&{7.10} & {16.55} & {16.60} & {0.43}
&{7.06} & {23.38} & {23.44} & {0.30}    
&\textbf{{7.79}} & \textbf{{15.21}} &\textbf{ {15.25}} & \textbf{{0.51}}  \\
2 
&{5.83} & {17.17} & {17.21} & {0.34}
&{3.63} & {16.05} & {16.08} & {0.23}
&{7.25} & {18.10} & {18.14} & {0.40} 
&{7.15} & {25.60} & {25.66} & {0.28}   
&\textbf{{7.74}} & \textbf{{15.08}} & \textbf{{15.11}} & \textbf{{0.51}} \\
3 
&{5.72} & {16.93} & {16.97} & {0.34} 
&{3.73} & {17.70} & {17.74} & {0.21}
&{6.84} & {18.20} & {18.24} & {0.38}
&{6.91} & {25.82} & {25.88} & {0.27}   
&\textbf{{7.50}} & \textbf{{16.65}} & \textbf{{16.68}} & \textbf{{0.45}}\\
4 
&{5.48} & {21.09} & {21.10} & {0.26}
&{3.58} & {18.72} & {18.72} & {0.19} 
&{6.84} & {29.69} & {29.70} & {0.23}
&{7.02} & {42.60} & {42.62} & {0.16}
&\textbf{{7.22}} & \textbf{{18.60}} & \textbf{{18.61}} & \textbf{{0.39}} \\
5 
&{6.55} & {20.40} & {20.44} & {0.32}
&{3.84} & {19.82} & {19.85} & {0.19}
&{7.56} & {22.61} & {22.65} & {0.33} 
&{7.33} & {31.92} & {31.98} & {0.23} 
&\textbf{{8.60}} & \textbf{{18.58}} & \textbf{{18.62}} & \textbf{{0.46}} \\
6 
&{6.90} & {21.27} & {21.34} & {0.32}
&{3.99} & {21.54} & {21.60} & {0.19} 
&{7.77} & {20.25} & {20.32} & {0.38} 
&{7.47} & {28.52} & {28.61} & {0.26} 
&\textbf{{8.98}} & \textbf{{19.06}} & \textbf{{19.12}} &\textbf{ {0.47}} \\
7 
&{5.67} & {21.79} & {21.80} & {0.26}
&{3.61} & {19.24} & {19.24} & {0.19} 
&{6.96} & {30.90} & {30.91} & {0.23}
&{7.02} & {43.58} & {43.59} & {0.16}
&\textbf{{7.45}} & \textbf{{19.20}} & \textbf{{19.20}} & \textbf{{0.39}} \\
8 
& {5.71} & {17.43} & {17.46} & {0.33}
&{3.65} & {16.71} & {16.74} & {0.22}
&{7.01} & {19.77} & {19.81} & {0.35}
&{6.99} & {27.46} & {27.51} & {0.25}
&\textbf{{7.54}} & \textbf{{15.33}} & \textbf{{15.36}} & \textbf{{0.49}} \\
9 
&{5.99} & {19.78} & {19.80} & {0.30} 
&{3.77} & {19.64} & {19.66} & {0.19} 
&{7.19} & {24.69} & {24.71} & {0.29}
&{7.11} & {34.52} & {34.55} & {0.21}  
&\textbf{{7.84}} & \textbf{{17.56}} & \textbf{{17.58}} & \textbf{{0.45}} \\
10 
&{5.73} & {17.74} & {17.78} & {0.32} 
&{3.69} & {17.77} & {17.81} & {0.21}
&{7.05} & {17.68} & {17.72} & {0.40}
&{6.98} & {24.44} & {24.49} & {0.29} 
&\textbf{{7.58}} & \textbf{{15.42}} & \textbf{{15.46}} & \textbf{{0.49}} \\

\Xhline{0.75pt}
{\textbf{Ave}} 
&5.95 & 19.02 & 19.06 & 0.32 
&3.72 & 18.37 & 18.39 &  0.21
&7.16 & 21.84 & 21.88 & 0.34
&7.10 & 30.78 & 30.83 & 0.24
&\textbf{7.82} & \textbf{17.07} & \textbf{17.10} & \textbf{0.46} \\
\Xhline{0.75pt}

\end{tabular}
}
\end{table*}

\begin{table*}[htbp]
\caption{Transfer learning on the BCI3 dataset.}
\label{tab: transfer learning}
    \centering
    \scalebox{0.83}{
\renewcommand{\arraystretch}{1.5}
\begin{tabular}{c:cccc:cccc:cccc:cccc:cccc}
\Xhline{0.75pt}
\multicolumn{1}{c}{\multirow{2}{*}{\textbf{Sub}}}
&\multicolumn{4}{c}{\textbf{ASAEDCT}~\cite{zhu2024electroencephalogram}} 
&\multicolumn{4}{c}{\textbf{KCHE}~\cite{idrees2022edge}}
&\multicolumn{4}{c}{\textbf{DCT-Perceptron}~\cite{pan2024multichannel}}
&\multicolumn{4}{c}{\textbf{VAE}~\cite{oliveira2023early}}
&\multicolumn{4}{c}{\textbf{AVDCT-Net}(Proposed)}\\
\Xcline{2-21}{0.75pt}
\multicolumn{1}{c}{\multirow{2}{*}{{}}}
&
\multicolumn{1}{c}{\textbf{CR}}&\multicolumn{1}{c}{\textbf{PRD}} &\multicolumn{1}{c}{\textbf{PRDN}} &\multicolumn{1}{c}{\textbf{QS}}  
&\multicolumn{1}{c}{\textbf{CR}}&\multicolumn{1}{c}{\textbf{PRD}} &\multicolumn{1}{c}{\textbf{PRDN}} &\multicolumn{1}{c}{\textbf{QS}}
&\multicolumn{1}{c}{\textbf{CR}}&\multicolumn{1}{c}{\textbf{PRD}} &\multicolumn{1}{c}{\textbf{PRDN}} &\multicolumn{1}{c}{\textbf{QS}}
&\multicolumn{1}{c}{\textbf{CR}}&\multicolumn{1}{c}{\textbf{PRD}} &\multicolumn{1}{c}{\textbf{PRDN}} &\multicolumn{1}{c}{\textbf{QS}}
&\multicolumn{1}{c}{\textbf{CR}}&\multicolumn{1}{c}{\textbf{PRD}} &\multicolumn{1}{c}{\textbf{PRDN}} &\multicolumn{1}{c}{\textbf{QS}}
 \\
\Xhline{0.75pt}
ATr 
&{11.12} & {9.73} & {9.75} & {1.14}  
&{6.80} & {8.78} & {8.80} & {0.77}
&{10.99} & {10.13} & {10.15} & {1.09}   
&{10.44} & {11.99} & {12.02} & {0.87}
&\textbf{{12.47}} & \textbf{{8.18}} & \textbf{{8.20}} & \textbf{{1.52}} \\
ATe 
&{10.08} & {12.13} & {12.16} & {0.83}
&{7.13} & {12.49} & {12.52} & {0.57}
&{10.64} & {13.11} & {13.14} & {0.81} 
&{10.32} & {15.89} & {15.92} & {0.65} 
&\textbf{{11.48}} & \textbf{{10.12}} & \textbf{{10.14}} & \textbf{{1.13}}  \\
BTr 
&{8.21} & {9.11} & {9.14} & {0.90} 
&{6.67} & {10.70} & {10.74} & {0.62}
&{10.08} & {10.77} & {10.81} & {0.94} 
&{9.73} & {11.06} & {11.10} & {0.88} 
&\textbf{{10.32}} & \textbf{{6.88}} & \textbf{{6.90}} & \textbf{{1.50}} \\
BTe 
&{10.49} & {9.87} & {9.88} & {1.06} 
&{6.50} & {8.46} & {8.47} & {0.77}
&{10.96} & {9.29} & {9.30} & {1.18} 
&{10.34} & {11.87} & {11.88} & {0.87} 
&\textbf{{11.88}} & \textbf{{7.77}} & \textbf{{7.78}} & \textbf{{1.53}} \\
\Xhline{0.75pt}
{\textbf{Ave}} 
&9.98 & 10.21 & 10.23 & 0.98
&6.78 & 10.11 & 10.13 & 0.68
&10.67 & 10.82 & 10.85 & 1.00
&10.21 & 12.70 & 12.73 & 0.82
&\textbf{11.54} & \textbf{8.24} & \textbf{8.26} & \textbf{1.42} \\
\Xhline{0.75pt}
\end{tabular}
}
\end{table*}

\subsection{Data Compression Experiment}
In the compression experiment on the BCI2 dataset, 70\% of the data from recording ``AA010'' was employed for model training.  The remaining 9 recordings (``AA001''--``AA009'') along with 30\% of recording ``AA010'' were utilized to test the model. In summary, the training set is composed of 1,885 samples, while the testing set comprises 25,120 samples. 
Besides, we choose $\tau=2$ and $\omega=1.2$ in Eq.~(\ref{Eq:code}).
Table~\ref{tab: Compression experiment on BCI2 dataset} summarizes the compression results of various models across the 10 subsets of the BCI2 dataset. 
In comparison to ASAEDCT, AVDCT-Net achieves a lower PRD with a higher CR, demonstrating its capacity to effectively balance compression efficiency with reconstruction accuracy.
The reason is that AVDCT-Net leverages a new ELBO as the loss function to train the model. This ELBO restricts the sum of the latent space coefficients to a relatively small, nonzero value instead of penalizing each coefficient towards 0. Therefore, AVDCT-Net can preserve the primary DCT coefficients while suppressing small high-frequency components. 
Besides, AVDCT-Net outperforms DCT-Perceptron because it applies linear layers and muti-head attention to enhance the capability to compress and restore the original signals.
Moreover, compared to KCHE, AVDCT-Net achieves a reduction in average PRD from 18.37 to 17.07 (a decrease of 7.08\%) and in PRDN from 18.39 to 17.10 (a decrease of 7.01\%). Furthermore, it enhances the CR from 3.72 to 7.82 (a 2.10-fold increase) and improves the QS from 0.21 to 0.46 (a 2.19-fold improvement).
 This is because AVDCT-Net incorporates multiple trainable hard-thresholding and scaling layers to the encoder, allowing it to eliminate redundant data more effectively than KCHE. 
 Additionally, AVDCT-Net is superior to VAE in terms of CR and PRD.
 While VAE integrates Gaussian processes to enhance feature extraction, it continues to utilize a conventional autoencoder structure. Its imperfect encoder and decoder result in higher reconstruction errors than AVDCT-Net.

 Table~\ref{tab: Compression experiment on BCI2 dataset} presents the compression results of various models initialized with a single random seed. To better assess model stability, we extend our experiment by initializing compression models with five different random seeds and computing the mean and standard deviation of CR, PRD, PRDN, and QS for each subset. The results are provided in Table I of the supplementary material. 
For subset 3, the CR is reported as 7.48 ± 0.04. The PRD is 16.12 ± 0.31, while the PRDN is 16.15 ± 0.31. Furthermore, the QS is measured at 0.46 ± 0.01. These results exhibit low variability, implying that AVDCT-Net has strong stability.
The reason is that each DCT compression unit utilizes a multi-channel architecture to capture features across different hierarchical levels and scales, facilitating a more comprehensive understanding of input EEG signals. This design mitigates dependency on specific initialization states and enhances the generalization capability.
Moreover, AVDCT-Net consistently outperforms ASAEDCT, KCHE, DCT-Perceptron, and VAE, achieving the best CR, PRD, PRDN, and QS across all 10 subsets, indicating its superior compression performance.

\subsection{Transfer Learning Experiment}
To assess the generative capabilities of various models, we perform a transfer learning experiment on the BCI3 dataset. In this experiment, the model trained on datasets obtained from Subject A is evaluated on datasets acquired from a different subject, Subject B. 
Specifically, 70\% of the data from Subject A's recording ATr was used for model training. The remaining 30\% of ATr, together with the recordings ATe, BTr, and BTe, were employed for model testing.
In brief, the training set is made up of 7,246 samples, and the testing sets comprise 3,106, 12,179, 10,352, and 12,179 samples, respectively.
Additionally, we select $\tau=4$ and $\omega=9.3$ in Eq.~(\ref{Eq:code}).
As tabulated in Table~\ref{tab: transfer learning}, AVDCT-Net demonstrates superior performance compared to other state-of-the-art approaches across all four subsets, with average CR improvements ranging from 8.15\% to 70.21\%, and average PRD enhancements varying between 18.50\% and 35.12\%. The reason is that ASAEDCT, KCHE, DCT-Perceptron, and VAE process each EEG channel independently, neglecting the inter-channel correlations. 
AVDCT-Net utilizes an adaptive filter bank, allowing each EEG channel to extract and integrate relevant features from adjacent channels, thereby mitigating the potential loss of critical information inherent in the compression process. Therefore, AVDCT-Net has a more robust generalization ability than other benchmark methods.

\subsection{Visual Assessment}
\begin{figure*}[htbp]
    \centering
    \subfloat[Comparison between the original data and the reconstructed data on the BCI2 dataset.]{
        \includegraphics[width=0.9\textwidth]{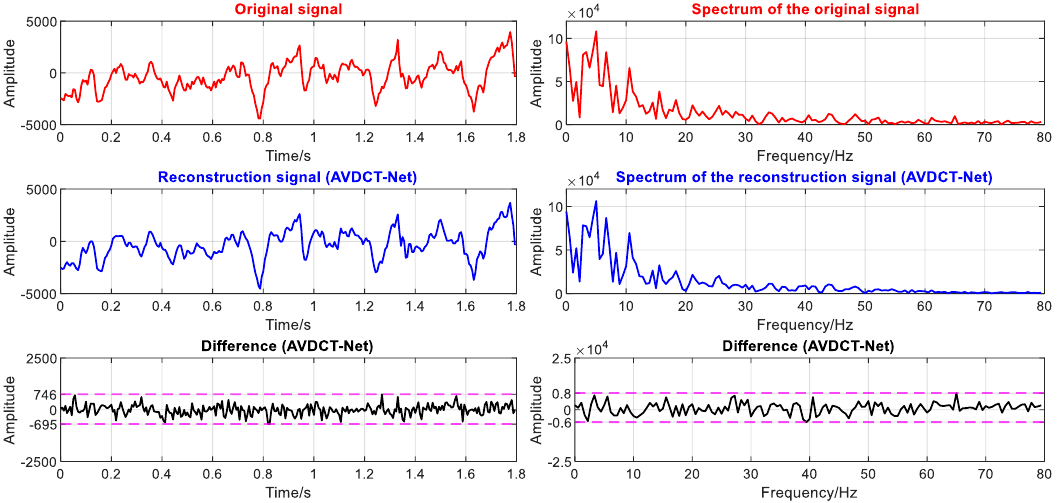}
        \label{fig:image1}
    }
    \\
    \subfloat[Comparison between the original data and the reconstructed data on the BCI3 dataset.]{
        \includegraphics[width=0.9\textwidth]{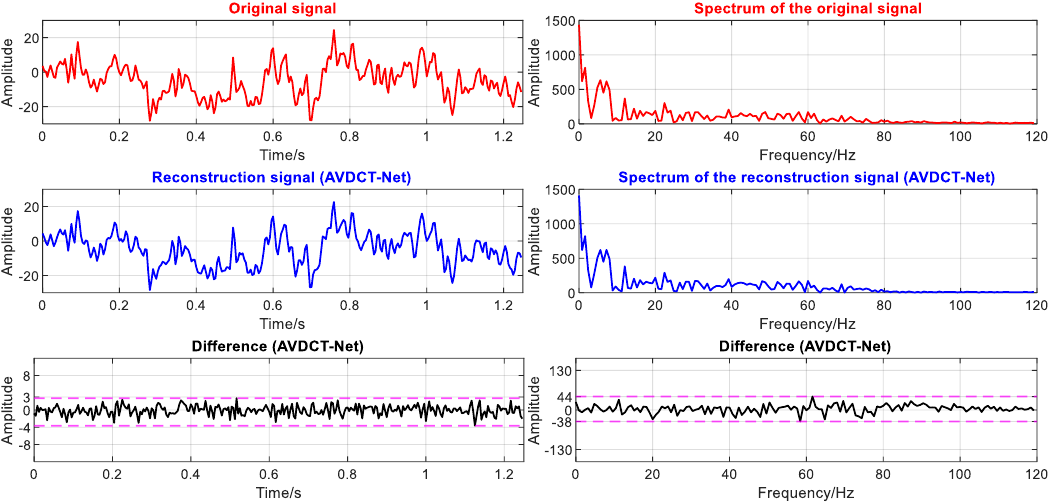}
        \label{fig:image2}
    }
    \caption{Comparison between the original data and the reconstructed data in the time domain (left) and the frequency domain (right) on the BCI2 dataset and BCI3 dataset.}
    \label{fig:overall}
\end{figure*}

To comprehensively evaluate the performance of the proposed compression algorithm, we present a visual comparison of the reconstructed EEG signals in both the time and frequency domains, as illustrated in Fig.~\ref{fig:overall}. In the time domain, it is observed that the differences between reconstructed and original signals are limited to a small range, indicating that AVDCT-Net effectively preserves the temporal features of the EEG data. In the frequency domain, the Fourier transform of the signals is analyzed. Results show that the frequency components of the reconstructed signal closely match those of the original signal. It demonstrates 
 AVDCT-Net effectively preserves the primary components in the low-frequency bands while also capturing key features in the high-frequency bands. 

AVDCT-Net encoder utilizes a hard-thresholding layer to eliminate small coefficients in the DCT domain, which predominantly correspond to high-frequency components. Although this process leads to information loss in the high-frequency band, the lost information consists of noise or redundant details. Therefore, their impact on signal reconstruction accuracy is minimal, as shown in Table~\ref{tab: Compression experiment on BCI2 dataset} and Table~\ref{tab: transfer learning}.
 Additionally, the AVDCT-Net decoder employs an adaptive filter bank to integrate salient features from adjacent channels into each EEG channel. While this mechanism enhances feature extraction, it introduces some errors in the low-frequency band. However, as depicted in Fig.~\ref{fig:overall}, these errors are limited to a small range. Consequently, their impact on reconstruction quality remains minimal. Besides, they do not compromise the performance of subsequent classification tasks as shown in Table~\ref{tab: P300 detection experiment}.


\subsection{Ablation Experiments }
\label{sec: Ablation experiments}
In this section, we perform ablation experiments to evaluate the impact of the primary components of the proposed method, including the scaling layer, MHA, adaptive filter bank, and ELBO. 
Additionally, we also investigated the effect of subband quantity on model compression performance. 
Moreover, CR, PRD and PRDN are three critical metrics for evaluating compression performance. Therefore, $\tau$ and $\omega$ in Eq.~(\ref{Eq:code}) are fine-tuned to optimize CR, PRD, and PRDN for each method, thereby demonstrating the effectiveness of the various modules. 
The results of the ablation experiments on the BCI2 dataset are summarized in Table~\ref{tab: Ablation study}.
\subsubsection{Ablation study on scaling layer}
Removing the scaling layer from the AVDCT-Net leads to an increase in the PRD from 17.07 to 17.28, representing a 1.23\% rise, and a corresponding decrease in the CR from 7.82 to 7.64, amounting to a 2.30\% reduction. This is due to the fact that the scaling layer in the DCT domain can be converted into the convolutional layer in the time domain. Consequently, it strengthens the encoder's ability to extract important features, leading to improved reconstruction accuracy without compromising the compression ratio.

\subsubsection{Ablation study on MHA}
In the absence of the MHA module within the decoder, a noticeable degradation in performance is observed, as reflected by a decreased CR, increased PRD, and a lower QS. 
MHA enables the simultaneous extraction of diverse features across different dimensions of the latent space. Additionally, it captures both local and global dependencies~\cite{guo2024temporal}, promoting better generalization and scalability in multi-channel EEG data reconstruction. Therefore, the MHA is essential in optimizing reconstruction accuracy.
\begin{table}[tb]
\caption{Ablation study.}
\label{tab: Ablation study}
    \centering
    \renewcommand{\arraystretch}{1.2}
    \begin{tabular*}{\linewidth}{@{\extracolsep{\fill}}lllll@{}}
\toprule
\multirow{2}{*}{\makecell{\textbf{Algorithm}}}&\textbf{Metrics}&  \\ \cmidrule[0.75pt]{2-5}
        \multirow{2}{*}{ } &\textbf{CR}
        &\textbf{PRD} &\textbf{PRDN}& \textbf{QS}  \\ 
\midrule
           No scaling layer&7.64 & 17.28 & 17.31 & 0.44 \\%
            No MHA &7.55 & 17.42 & 17.45 & 0.44\\%
           No adaptive filter bank &7.69 & 17.26 & 17.30 & 0.45 \\%
            No ELBO &7.64 & 17.17 & 17.20 & 0.45  \\   %
           {AVDCT-Net  (Channel=1)} &7.56 & 17.24 & 17.27 & 0.44  \\%
          {AVDCT-Net  (Channel=2)} &7.69 & 17.24 & 17.27 & 0.45 \\ %
            \textbf{AVDCT-Net  (Channel=3)} &\textbf{7.82} & \textbf{17.07} & \textbf{17.10} & \textbf{0.46}  \\%
            {AVDCT-Net  (Channel=4)} & 7.81 & 17.59 & 17.62 & 0.45  \\%
            \bottomrule
\end{tabular*}
\vskip -0.1in
\end{table}
\subsubsection{Ablation study on adaptive filter bank}
When the adaptive filter bank is not present in the AVDCT-Net, the CR is decreased from 7.82 to 7.69 (1.66\%), and the PRD is increased from 17.07 to 17.26 (1.11\%). This is because an adaptive filter bank allows each EEG channel to merge important features from neighbouring channels. This helps restore critical information that may be lost during the lossy compression procedure. Thus, the adaptive filter bank plays a crucial role in enhancing the accuracy of data reconstruction.

\subsubsection{Ablation study on ELBO}
Compared to AVDCT-Net trained with the MSE loss function, AVDCT-Net optimized using the ELBO loss function exhibits an improved compression quality score as shown in Table~\ref{tab: Ablation study}. By balancing the trade-off between the reconstruction loss and the KL divergence term, ELBO inherently penalizes the model for introducing unnecessary complexity in the latent space. If certain latent variables do not contribute to improving the reconstruction accuracy, the KL divergence term drives the model to eliminate these variables by constraining their distribution to match the predefined distribution. This process removes redundant coefficients, enhancing the compression efficiency.

\subsubsection{Ablation study on subband quantity}
Compared to AVDCT-Net with a single subband, AVDCT-Net with three subbands achieves a 3.44\% improvement in CR, a 0.99\% improvement in PRD, a 0.98\% improvement in PRDN, and a 4.55\% improvement in QS. The reason is that three scaling subbands in the DCT domain can be treated as three convolution layers in the time domain. Their coordinated operation synthesizes a filter bank that improves feature extraction efficiency. Furthermore, compared with other multiple subband models, the three-subband model exhibits the highest CR along with the lowest PRD and PRDN. Therefore, we select a subband quantity of three.

\subsection{Computational Complexity and Compression Time}

Table~\ref{table: MACs} provides a comparative analysis of the trainable parameters and Multiply-Accumulate Operations (MACs) within the encoder modules across various algorithms. 
We focus on analyzing the encoder module because it is deployed in a resource-constrained edge gateway, whereas the decoder can be implemented on a more capable fog
gateway.
In this study, we leverage a 64-channel EEG signal as input, with each channel comprising 64 data samples captured over a duration of 0.27 seconds. 
Compared to the VAE, AVDCT-Net reduces the number of trainable parameters from 11,700 to 4,547, representing a 61.14\% reduction. This significantly lowers the hardware memory requirements for parameter storage. Furthermore, AVDCT-Net decreases MACs from 737,280 to 561,152, a reduction of 23.89\%, indicating a lower computational complexity.
Additionally, AVDCT-Net has a comparable number of trainable parameters and MACs to ASAEDCT.

To evaluate the practicality of the proposed compression model in real-world scenarios, we analyze its actual energy consumption and inference latency on edge devices. For this study, the Raspberry Pi 400~\cite{RaspberryPi400,mohammad2024iot} is chosen as the edge computing platform. The Raspberry Pi 400 features a Broadcom BCM2711 system-on-chip (SoC) with a quad-core Cortex-A72 (ARM v8) 64-bit processor running at 1.8GHz. It is equipped with 4GB of LPDDR4-3200 SDRAM, enabling efficient multitasking and computing capabilities. 
Additionally, the primary energy consumption on the Raspberry Pi 400 occurs during the data compression process. It is computed as $E = P \times t$, where $P$ is the power in watts (W), measured directly using a power meter, and $t$ is the inference time in seconds (s). 
As shown in Table~\ref{table: Power}, AVDCT-Net demonstrates a lower inference latency (0.0067s) and power cost (0.0422W) compared to ASAEDCT and VAE. The reason is that ASAEDCT employs an additional Tanh activation function compared to AVDCT-Net, while VAE utilizes more linear layers than AVDCT-Net.
Although KCHE requires no trainable parameters and has fewer MACs per iteration, it exhibits higher inference latency and power consumption than AVDCT-Net. This is because KCHE demands numerous iterations to achieve convergence. In contrast, AVDCT-Net requires only a single forward pass during the inference stage. Furthermore, AVDCT-Net consumes 21.61\% more energy and incurs 21.82\% higher latency than DCTP. However, it improves compression quality by 35.29\%, as shown in Table~\ref{tab: Compression experiment on BCI2 dataset}. This trade-off highlights AVDCT-Net’s ability to balance compression performance and power efficiency, making it well-suited for deployment on edge devices.


\begin{table}[t]
\centering
\caption{The comparison of the number of trainable parameters and MACs.}\label{table: MACs}
\renewcommand{\arraystretch}{1.2}
\begin{tabular}{lccc}
\toprule
\multirow{2}{*}{\makecell{\textbf{Algorithm}}}   & \multirow{2}{*}{\makecell{\textbf{Input size}}}  & \textbf{Trainable} &\multirow{2}{*}{\makecell{\textbf{MACs}}}\\ 
{}& {}& \textbf{parameters} &{}\\
\midrule
ASAEDCT~\cite{zhu2024electroencephalogram}         & 64$\times$64   & 4,547   &561,152  \\
KCHE~\cite{idrees2022edge}            & 64$\times$64   & 0       & 12,352    \\
DCT-Perceptron~\cite{pan2024multichannel}  & 64$\times$64   & 129     &274,432 \\
VAE~\cite{oliveira2023early}             & 64$\times$64   & 11,700  &737,280 \\
\textbf{AVDCT-Net}& 64$\times$64   &4,547    &561,152 \\
\bottomrule
\end{tabular}
\end{table}


\begin{table}[t]
\centering
\caption{The comparison of the power cost and runtime on Raspberry Pi 400.}\label{table: Power}
\renewcommand{\arraystretch}{1.2}
\begin{tabular}{lccc}
\toprule
\multirow{2}{*}{\makecell{\textbf{Algorithm}}}   &\multirow{2}{*}{\makecell{\textbf{Input size}}}  & \textbf{Power cost}& \textbf{Runtime}\\ 
{}&{} & \textbf{(Joule)} & {\textbf{(s)}}\\
\midrule
ASAEDCT~\cite{zhu2024electroencephalogram}         & 64$\times$64   &0.0510    &0.0081 \\
KCHE~\cite{idrees2022edge}            & 64$\times$64   & 0.2904        & 0.0484    \\
DCT-Perceptron~\cite{pan2024multichannel}  & 64$\times$64   &0.0347    &0.0055\\
VAE~\cite{oliveira2023early}             & 64$\times$64   &0.0617  &0.0098 \\
\textbf{AVDCT-Net}& 64$\times$64   &0.0422     &0.0067\\
\bottomrule
\end{tabular}
\end{table}

\begin{figure}[t]
	\centering
		\includegraphics[scale=.4]{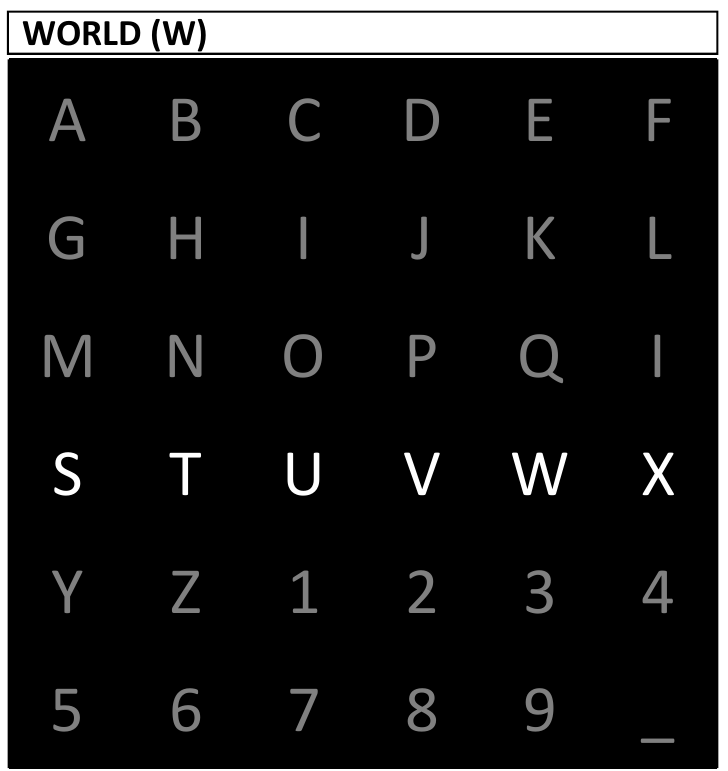}
	\caption{A 6x6 P300 speller. In this example, the user is required to spell the word ``WORLD” (one character at a time). For each character, the application randomly intensifies columns and rows several times (e.g., the fourth row in this instance)}
	\label{Fig:P300}
\end{figure}

\subsection{P300 Detection Experiment}
The EEG data compression model is developed for remote BCI systems. Therefore, the P300 detection experiment is conducted to validate the effectiveness of the compression model. In the P300 detection experiment~\cite{blankertz2006bci}, the participants are shown a $6\times 6$ matrix composed of 36 individual alphanumeric characters as shown in Fig.~\ref{Fig:P300}. The task of the participants is to sequentially focus their gaze on lighted characters that are predetermined by the researcher. Each row and column of the $6\times 6$ matrix was randomly intensified, yielding 12 distinct stimuli—6 corresponding to the rows and 6 to the columns. Among these intensifications, two specifically highlighted the target character. In this process, the brain activities of participants evoked by two stimuli with the target character differ from those evoked by the stimuli without the target character. Therefore, according to such differences, the BCI system can predict the character that participants were focusing on.
The detailed experimental content is described in~\cite{blankertz2006bci}.
In this section, we utilize the BCI3 dataset because it provides a comprehensive record of P300 evoked potentials captured using the BCI2000 software. The paradigm utilized for these recordings is outlined in [13]. 
In the BCI3 datasets, the training sets contain recordings of 85 distinct characters, while the testing sets include recordings of 100 different characters. 
Each character is labeled as either a target or non-target. 
In addition, a single channel records 7,560 data points for each character.

\begin{figure*}[t]
	\centering
		\includegraphics[scale=.18]{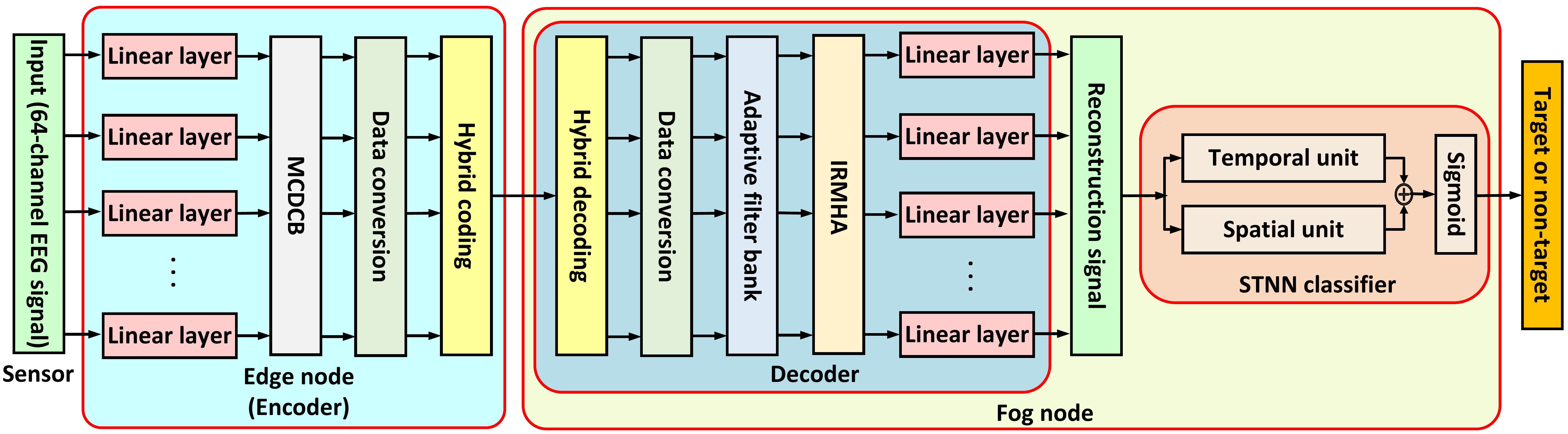}
	\caption{ P300 detection model consisting of the data compression/decompression and STNN network.}
	\label{Fig:compress+classfication}
\end{figure*}

\begin{figure*}[b]
	\centering
		\includegraphics[scale=.62]{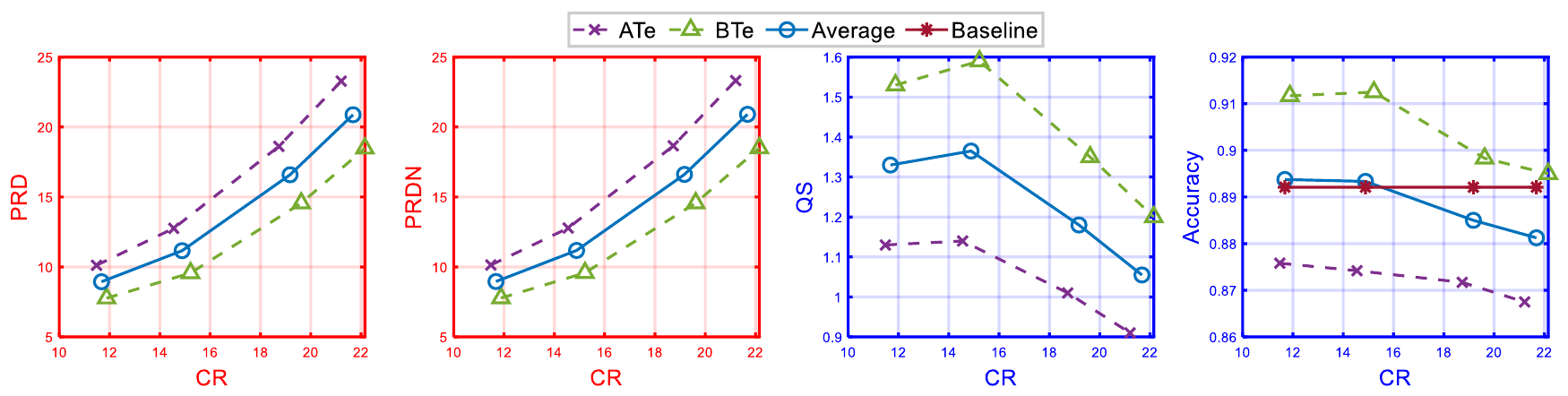}
	\caption{Compression performance of AVDCT-Net and classification accuracy of STNN for different CRs.}
	\label{Fig:cr-acc}
\end{figure*}

The experiment begins with the application of the pre-trained AVDCT-Net encoder to compress the EEG data at the edge node as shown in Fig.~\ref{Fig:compress+classfication}. The compressed data is then converted into bitstreams for wireless transmission. Subsequently, the pre-trained AVDCT-Net decoder is utilized to restore the original EEG data at the fog node. After that, a spatial-temporal neural network (STNN)~\cite{zhang2021spatial} is employed for P300 detection. The STNN has demonstrated superior performance across diverse P300 detection tasks. It is composed of a temporal unit and a spatial unit. 
The detailed structure of the STNN is illustrated in the supplementary material. Given that the authors in~\cite{zhang2021spatial} also employed STNN for P300 detection on the BCI3 dataset, we adhere to the same data preprocessing, training, and testing procedures as described in their research. Specifically, the original recordings ATr and BTr are utilized as the training datasets, while the recordings ATe and BTe, reconstructed through compression models, are employed as the testing datasets.

\begin{table}[t]
\caption{P300 detection experiment on the BCI3 dataset.}
\label{tab: P300 detection experiment}
    \centering
    \renewcommand{\arraystretch}{1.2}
    \begin{tabular*}{\linewidth}{@{\extracolsep{\fill}}lccc@{}}
\toprule
\multirow{2}{*}{\makecell{\textbf{Algorithm}}}&\textbf{Accuracy}&  \\ \cmidrule[0.75pt]{2-4}
        \multirow{2}{*}{ } &\textbf{Subject-A}
        &\textbf{Subject-B} &\textbf{Ave}  \\ 
\midrule
    Original+STNN       &{87.50\%}    & {90.92\%}   & {89.21\%}  \\
    ASAEDCT+STNN        &{87.58\%}    & {90.83\%}   & {89.21\%}  \\
    KCHE+STNN           &{87.00\%}   & {90.58\%}   & {88.79\%}  \\
    DCT-Perceptron+STNN &{87.00\%}   & {91.00\%}   & {89.00\%}   \\
    VAE+STNN            &{86.92\%}     & {90.92\%}   & {88.92\%}   \\   
    \textbf{AVDCT-Net+STNN }        &{\textbf{87.58\%}}   & {\textbf{91.17\%}}   & {\textbf{89.38\%}}  \\
            \bottomrule
\end{tabular*}
 \vskip -0.1in
\end{table}

As shown in Table~\ref{tab: P300 detection experiment}, the STNN classifier achieved higher accuracy on datasets reconstructed by AVDCT-Net compared to the original datasets. It indicates that AVDCT-Net effectively preserves key classfication features during data compression and reconstruction. Additionally, AVDCT-Net is capable of eliminating redundant high-frequency components, such as noise, from EEG signals,  which contributes to the improvement in accuracy.
 Furthermore, among datasets reconstructed by various compression models, the STNN classifier attains the highest average accuracy rate of 89.38\% on those reconstructed by AVDCT-Net.
 This demonstrates that AVDCT-Net retains more important information than other models during the compression process.
During compression, AVDCT-Net employs two key steps to preserve and enhance critical classification features in the DCT domain. First, a hard-thresholding layer filters out small DCT coefficients, which mainly represent noise and redundant details, while retaining large coefficients that hold vital classification information. Second, the scaling layer works as a trainable filter to further emphasize important features. As shown in Fig.~\ref{fig:overall}, AVDCT-Net effectively suppresses redundancy by discarding small coefficients in the frequency domain while preserving significant ones. Additionally, STNN classifies EEG signals by leveraging both temporal and global features. Meanwhile, the IRMHA mechanism in AVDCT-Net effectively captures a wide range of local and global dependencies from the feature map, thereby amplifying classification features.

Fig.~\ref{Fig:cr-acc} shows the impact of different CR on the compression performance of AVDCT-Net and the accuracy rate of the STNN classifier, evaluated on dataset ATe and dataset BTe. In this experiment, the adjustment of CR is achieved by modifying the values of $\tau$ and $\omega$ as defined in Eq. (\ref{Eq:code}). As the average CR increases from 11.68, both PRD and PRDN gradually rise, while the average accuracy of the STNN classifier on datasets reconstructed by AVDCT-Net gradually declines. This decline occurs because, with higher CR, more important information is lost from the signal. Additionally, it is observed that when the average CR is 14.89, the classifier’s average accuracy aligns with its performance on original datasets (baseline). This implies that the average CR can be increased to at least 14.89 without compromising average classification accuracy. Moreover, AAEVDCT attains its highest QS at a CR of 14.89, indicating an optimal tradeoff between compression efficiency and reconstruction accuracy.

\section{Conclusion}
In this article, we proposed a novel AVDCT-Net model for EEG data compression within an edge-fog computing architecture. The model leveraged a multi-channel structure to enable the concurrent compression of EEG signals across all channels, reducing both compression time and transmission latency.
Specifically, a low-complexity encoder was deployed at the edge level. By imposing parallel hard-thresholding nonlinearities and scaling operators (filters) after the DCT, the redundant coefficients were removed, thereby enhancing the encoder's data compression capabilities. 
Besides, a new ELBO was derived to regularize the encoder output, promoting a more structured and efficient representation.
At the fog level, an adaptive filter bank was adopted to merge relevant features from adjacent channels into each separate channel, improving the robustness and generalization ability. Furthermore, the IRMHA captured both local and global dependencies to reconstruct the original signal. 
The proposed model exhibited superior compression efficiency and reconstruction accuracy compared to state-of-the-art methods on the BCI2 and BCI3 datasets. Furthermore, as validated by the P300 detection experiment, AVDCT-Net preserved all critical features during the compression and reconstruction process. Therefore, the AVDCT-Net model was well-suited for implementation in an edge-fog computing framework for EEG data compression. 
In future work, AVDCT-Net will be deployed across a wider range of edge devices, including low-power IoT devices (e.g., Raspberry Pi, Arduino), embedded systems (e.g., FPGAs, DSPs), mobile devices (e.g., smartphones, tablets), and industrial/automotive edge computing platforms (e.g., NVIDIA Jetson, Qualcomm Snapdragon).

\section*{Acknowledgments}
This work was supported by the National Science Foundation (NSF) IDEAL 2217023.

\bibliographystyle{IEEEbib}
\bibliography{strings}

\begin{thebibliography}{10}

\bibitem{douibi2021toward}
Khalida Douibi, Sol{\`e}ne Le~Bars, Alice Lemontey, Lipsa Nag, Rodrigo Balp, and Gabri{\`e}le Breda,
\newblock ``Toward eeg-based bci applications for industry 4.0: Challenges and possible applications,''
\newblock {\em Frontiers in Human Neuroscience}, vol. 15, pp. 705064, 2021.

\bibitem{maiseli2023brain}
Baraka Maiseli, Abdi~T Abdalla, Libe~V Massawe, Mercy Mbise, Khadija Mkocha, Nassor~Ally Nassor, Moses Ismail, James Michael, and Samwel Kimambo,
\newblock ``Brain--computer interface: trend, challenges, and threats,''
\newblock {\em Brain informatics}, vol. 10, no. 1, pp. 20, 2023.

\bibitem{an2024development}
Yang An, Johnny Wong, and Sai~Ho Ling,
\newblock ``Development of real-time brain-computer interface control system for robot,''
\newblock {\em Applied Soft Computing}, vol. 159, pp. 111648, 2024.

\bibitem{bagheri2022simultaneous}
Mahsa Bagheri and Sarah~D Power,
\newblock ``Simultaneous classification of both mental workload and stress level suitable for an online passive brain--computer interface,''
\newblock {\em Sensors}, vol. 22, no. 2, pp. 535, 2022.

\bibitem{khan2021analysis}
Haroon Khan, Noman Naseer, Anis Yazidi, Per~Kristian Eide, Hafiz~Wajahat Hassan, and Peyman Mirtaheri,
\newblock ``Analysis of human gait using hybrid eeg-fnirs-based bci system: a review,''
\newblock {\em Frontiers in Human Neuroscience}, vol. 14, pp. 613254, 2021.

\bibitem{altaheri2022physics}
Hamdi Altaheri, Ghulam Muhammad, and Mansour Alsulaiman,
\newblock ``Physics-informed attention temporal convolutional network for eeg-based motor imagery classification,''
\newblock {\em IEEE transactions on industrial informatics}, vol. 19, no. 2, pp. 2249--2258, 2022.

\bibitem{rodriguez2023fog}
Paula~Ivone Rodr{\'\i}guez-Azar, Jose~Manuel Mej{\'\i}a-Mu{\~n}oz, Oliverio Cruz-Mej{\'\i}a, Rafael Torres-Escobar, and Lucero Ver{\'o}nica~Ruelas L{\'o}pez,
\newblock ``Fog computing for control of cyber-physical systems in industry using bci,''
\newblock {\em Sensors}, vol. 24, no. 1, pp. 149, 2023.

\bibitem{paszkiel2020using}
Szczepan Paszkiel and Szczepan Paszkiel,
\newblock ``Using bci in iot implementation,''
\newblock {\em Analysis and Classification of EEG Signals for Brain--Computer Interfaces}, pp. 111--128, 2020.

\bibitem{idrees2022edge}
Ali~Kadhum Idrees, Sara~Kadhum Idrees, Raphael Couturier, and Tara Ali-Yahiya,
\newblock ``An edge-fog computing-enabled lossless eeg data compression with epileptic seizure detection in iomt networks,''
\newblock {\em IEEE Internet of Things Journal}, vol. 9, no. 15, pp. 13327--13337, 2022.

\bibitem{hurbungs2021fog}
V~Hurbungs, V~Bassoo, and TP~Fowdur,
\newblock ``Fog and edge computing: concepts, tools and focus areas,''
\newblock {\em International Journal of Information Technology}, vol. 13, no. 2, pp. 511--522, 2021.

\bibitem{zhu2024electroencephalogram}
Xin Zhu, Hongyi Pan, Shuaiang Rong, and Ahmet~Enis Cetin,
\newblock ``Electroencephalogram sensor data compression using an asymmetrical sparse autoencoder with a discrete cosine transform layer,''
\newblock in {\em ICASSP 2024-2024 IEEE International Conference on Acoustics, Speech and Signal Processing (ICASSP)}. IEEE, 2024, pp. 2160--2164.

\bibitem{birvinskas2015fast}
Darius Birvinskas, Vacius Jusas, Ignas Martisius, and Robertas Damasevicius,
\newblock ``Fast dct algorithms for eeg data compression in embedded systems,''
\newblock {\em Computer Science and Information Systems}, vol. 12, no. 1, pp. 49--62, 2015.

\bibitem{nguyen2018study}
Binh Nguyen, Wanli Ma, and Dat Tran,
\newblock ``A study of combined lossy compression and seizure detection on epileptic {EEG} signals,''
\newblock {\em Procedia Computer Science}, vol. 126, pp. 156--165, 2018.

\bibitem{hejrati2017efficient}
Behzad Hejrati, Abdolhossein Fathi, and Fardin Abdali-Mohammadi,
\newblock ``Efficient lossless multi-channel eeg compression based on channel clustering,''
\newblock {\em Biomedical Signal Processing and Control}, vol. 31, pp. 295--300, 2017.

\bibitem{al2018convolutional}
Abeer~Z Al-Marridi, Amr Mohamed, and Aiman Erbad,
\newblock ``Convolutional autoencoder approach for {EEG} compression and reconstruction in m-health systems,''
\newblock in {\em 2018 14th International Wireless Communications \& Mobile Computing Conference (IWCMC)}. IEEE, 2018, pp. 370--375.

\bibitem{zancanaro2023veegnet}
Alberto Zancanaro, Giulia Cisotto, Italo Zoppis, and Sara~Lucia Manzoni,
\newblock ``veegnet: learning latent representations to reconstruct eeg raw data via variational autoencoders,''
\newblock in {\em International Conference on Information and Communication Technologies for Ageing Well and e-Health}. Springer, 2023, pp. 114--129.

\bibitem{lerogeron2023learning}
Hugo Lerogeron, Romain Picot-Cl{\'e}mente, Laurent Heutte, and Alain Rakotomamonjy,
\newblock ``Learning an autoencoder to compress {EEG} signals via a neural network based approximation of {DTW},''
\newblock {\em Procedia Computer Science}, vol. 222, pp. 448--457, 2023.

\bibitem{hejrati2017new}
Behzad Hejrati, Abdolhossein Fathi, and Fardin Abdali-Mohammadi,
\newblock ``A new near-lossless {EEG} compression method using {ANN}-based reconstruction technique,''
\newblock {\em Computers in Biology and Medicine}, vol. 87, pp. 87--94, 2017.

\bibitem{blankertz2004bci}
Benjamin Blankertz, K-R Muller, Gabriel Curio, Theresa~M Vaughan, Gerwin Schalk, Jonathan~R Wolpaw, Alois Schlogl, Christa Neuper, Gert Pfurtscheller, Thilo Hinterberger, et~al.,
\newblock ``The bci competition 2003: progress and perspectives in detection and discrimination of eeg single trials,''
\newblock {\em IEEE transactions on biomedical engineering}, vol. 51, no. 6, pp. 1044--1051, 2004.

\bibitem{blankertz2006bci}
Benjamin Blankertz, K-R Muller, Dean~J Krusienski, Gerwin Schalk, Jonathan~R Wolpaw, Alois Schlogl, Gert Pfurtscheller, Jd~R Millan, Michael Schroder, and Niels Birbaumer,
\newblock ``The bci competition iii: Validating alternative approaches to actual bci problems,''
\newblock {\em IEEE transactions on neural systems and rehabilitation engineering}, vol. 14, no. 2, pp. 153--159, 2006.

\bibitem{ahmed1974discrete}
Nasir Ahmed, T\_ Natarajan, and Kamisetty~R Rao,
\newblock ``Discrete cosine transform,''
\newblock {\em IEEE Transactions on Computers}, vol. 100, no. 1, pp. 90--93, 1974.

\bibitem{wallace1992jpeg}
Gregory~K Wallace,
\newblock ``The {JPEG} still picture compression standard,''
\newblock {\em IEEE Transactions on Consumer Electronics}, vol. 38, no. 1, pp. xviii--xxxiv, 1992.

\bibitem{le1991mpeg}
Didier Le~Gall,
\newblock ``Mpeg: A video compression standard for multimedia applications,''
\newblock {\em Communications of the ACM}, vol. 34, no. 4, pp. 46--58, 1991.

\bibitem{chamain2022end}
Lahiru~D Chamain, Siyu Qi, and Zhi Ding,
\newblock ``End-to-end image classification and compression with variational autoencoders,''
\newblock {\em IEEE Internet of Things Journal}, vol. 9, no. 21, pp. 21916--21931, 2022.

\bibitem{arican2019pairwise}
Murat Arican and Kemal Polat,
\newblock ``Pairwise and variance based signal compression algorithm (pvbsc) in the p300 based speller systems using eeg signals,''
\newblock {\em Computer methods and programs in biomedicine}, vol. 176, pp. 149--157, 2019.

\bibitem{donoho1995noising}
David~L Donoho,
\newblock ``De-noising by soft-thresholding,''
\newblock {\em IEEE Transactions on Information Theory}, vol. 41, no. 3, pp. 613--627, 1995.

\bibitem{martucci1994symmetric}
Stephen~A Martucci,
\newblock ``Symmetric convolution and the discrete sine and cosine transforms,''
\newblock {\em IEEE Transactions on Signal Processing}, vol. 42, no. 5, pp. 1038--1051, 1994.

\bibitem{park2003m}
HyunWook Park, YoungSeo Park, and Seung-Kyun Oh,
\newblock ``L/{M}-fold image resizing in block-{DCT} domain using symmetric convolution,''
\newblock {\em IEEE Transactions on Image Processing}, vol. 12, no. 9, pp. 1016--1034, 2003.

\bibitem{dauwels2012near}
Justin Dauwels, K~Srinivasan, M~Ramasubba Reddy, and Andrzej Cichocki,
\newblock ``Near-lossless multichannel eeg compression based on matrix and tensor decompositions,''
\newblock {\em IEEE journal of biomedical and health informatics}, vol. 17, no. 3, pp. 708--714, 2012.

\bibitem{zhu2024novel}
Xin Zhu, Daoguang Yang, Hongyi Pan, Hamid~Reza Karimi, Didem Ozevin, and Ahmet~Enis Cetin,
\newblock ``A novel asymmetrical autoencoder with a sparsifying discrete cosine stockwell transform layer for gearbox sensor data compression,''
\newblock {\em Engineering Applications of Artificial Intelligence}, vol. 127, pp. 107322, 2024.

\bibitem{ng2011sparse}
Andrew Ng et~al.,
\newblock ``Sparse autoencoder,''
\newblock {\em CS294A Lecture Notes}, vol. 72, no. 2011, pp. 1--19, 2011.

\bibitem{lam2000mathematical}
Edmund~Y Lam and Joseph~W Goodman,
\newblock ``A mathematical analysis of the dct coefficient distributions for images,''
\newblock {\em IEEE transactions on image processing}, vol. 9, no. 10, pp. 1661--1666, 2000.

\bibitem{akhter2010ecg}
Shahin Akhter and MA~Haque,
\newblock ``{ECG} compression using run length encoding,''
\newblock in {\em 2010 18th European Signal Processing Conference}. IEEE, 2010, pp. 1645--1649.

\bibitem{tu2006novel}
Zongjie Tu and Shiyong Zhang,
\newblock ``A novel implementation of {JPEG} 2000 lossless coding based on {LZMA},''
\newblock in {\em The Sixth IEEE International Conference on Computer and Information Technology (CIT'06)}. IEEE, 2006, pp. 140--140.

\bibitem{ratanaworabhan2006fast}
Paruj Ratanaworabhan, Jian Ke, and Martin Burtscher,
\newblock ``Fast lossless compression of scientific floating-point data,''
\newblock in {\em Data Compression Conference (DCC'06)}. IEEE, 2006, pp. 133--142.

\bibitem{pan2024multichannel}
Hongyi Pan, Emadeldeen Hamdan, Xin Zhu, Salih Atici, and Ahmet~Enis Cetin,
\newblock ``Multichannel orthogonal transform-based perceptron layers for efficient resnets,''
\newblock {\em IEEE Transactions on Neural Networks and Learning Systems}, 2024.

\bibitem{oliveira2023early}
Adaiton Oliveira-Filho, Ryad Zemouri, Philippe Cambron, and Antoine Tahan,
\newblock ``Early detection and diagnosis of wind turbine abnormal conditions using an interpretable supervised variational autoencoder model,''
\newblock {\em Energies}, vol. 16, no. 12, pp. 4544, 2023.

\bibitem{loshchilov2017decoupled}
Ilya Loshchilov and Frank Hutter,
\newblock ``Decoupled weight decay regularization,''
\newblock {\em arXiv preprint arXiv:1711.05101}, 2017.

\bibitem{christianos2021scaling}
Filippos Christianos, Georgios Papoudakis, Muhammad~A Rahman, and Stefano~V Albrecht,
\newblock ``Scaling multi-agent reinforcement learning with selective parameter sharing,''
\newblock in {\em International Conference on Machine Learning}. PMLR, 2021, pp. 1989--1998.

\bibitem{guo2024temporal}
Baosu Guo, Zhaohui Qiao, Hao Dong, Zhen Wang, Shuiquan Huang, Zhengkai Xu, Fenghe Wu, Chuanzhen Huang, and Qing Ni,
\newblock ``Temporal convolutional approach with residual multi-head attention mechanism for remaining useful life of manufacturing tools,''
\newblock {\em Engineering Applications of Artificial Intelligence}, vol. 128, pp. 107538, 2024.

\bibitem{RaspberryPi400}
{Raspberry Pi Foundation},
\newblock ``{Raspberry Pi 400 Product Brief},'' 2022,
\newblock Accessed: Feb. 16, 2025.

\bibitem{mohammad2024iot}
Ahmad~Saeed Mohammad, Thoalfeqar~G Jarullah, Musab~TS Al-Kaltakchi, Jabir Alshehabi Al-Ani, and Somdip Dey,
\newblock ``Iot-mfacenet: Internet-of-things-based face recognition using mobilenetv2 and facenet deep-learning implementations on a raspberry pi-400,''
\newblock {\em Journal of Low Power Electronics and Applications}, vol. 14, no. 3, pp. 46, 2024.

\bibitem{zhang2021spatial}
Zhen Zhang, Xiaoyan Yu, Xianwei Rong, and Makoto Iwata,
\newblock ``Spatial-temporal neural network for p300 detection,''
\newblock {\em IEEE Access}, vol. 9, pp. 163441--163455, 2021.

\end{thebibliography}

\vfill

\end{document}